\documentclass[onecolumn,amsmath,amssymb,12pt,superscriptaddress,nofootinbib]{revtex5}
\pdfoutput=1

\usepackage[latin1]{inputenc}
\usepackage[english]{babel}
\usepackage{amssymb}
\usepackage{amsmath}
\usepackage{amsthm}
\usepackage[]{graphicx}
\usepackage[]{subfigure}
\usepackage{tensor}
\usepackage{color}
\usepackage{cancel}
\usepackage{setspace}
\usepackage{fancyhdr}
\usepackage[bookmarks,linktocpage, colorlinks=true, plainpages = false, citecolor = blue,  linkcolor=blue, urlcolor = blue, filecolor = blue]{hyperref} 
\def\be{\begin{equation}}
\def\ee{\end{equation}}
\def\d{\mathrm{d}}

\def\cH{{\cal H}}

\def\p{\partial}
\def\d{\mathrm{d}}
\def\MP{\mathrm{M}_\mathrm{P}}
\def\psib{{\bf \psi}}
\def\Phib{{\bf \Phi}}

\usepackage{color}

\begin{document}

\allowdisplaybreaks

\begin{titlepage}

\title{Non--Singular Bouncing Cosmology:\\
Consistency of the Effective Description}

\author{Michael Koehn}
\email[]{koehn@physics.upenn.edu}
\affiliation{Department of Physics, University of Pennsylvania,\\ 209 South 33rd Street, Philadelphia, PA 19104-6395, U.S.A.}
\author{Jean-Luc Lehners}
\email[]{jlehners@aei.mpg.de}
\affiliation{Max Planck Institute for Gravitational Physics (Albert Einstein Institute), 14476 Potsdam, Germany}
\author{Burt Ovrut}
\email[]{ovrut@elcapitan.hep.upenn.edu}
\affiliation{Department of Physics, University of Pennsylvania,\\ 209 South 33rd Street, Philadelphia, PA 19104-6395, U.S.A.}

\begin{abstract}

\vspace{.3in}
\noindent 
We explicitly confirm that spatially flat non--singular bouncing cosmologies make sense as effective theories. The presence of a non-singular bounce in a spatially flat universe implies a temporary violation of the null energy condition, which can be achieved through a phase of ghost condensation. We calculate the scale of strong coupling and demonstrate that the ghost--condensate bounce remains trustworthy throughout, and that all perturbation modes within the regime of validity of the effective description remain under control. For this purpose we require the perturbed action up to third order in perturbations, which we calculate in both flat and co-moving gauge--since these two gauges allow us to highlight different physical aspects. Our conclusion is that there exist healthy descriptions of non--singular bouncing cosmologies providing a viable resolution of the big--bang singularities in cosmological models. Our results also suggest a variant of ekpyrotic cosmology, in which entropy perturbations are generated during the contracting phase, but are only converted into curvature perturbations after the bounce.
\end{abstract}
\maketitle

\end{titlepage}

\tableofcontents

\section{Introduction}

Almost a hundred years ago the discovery that the universe is expanding brought about a major paradigm shift in cosmological thinking: the universe is not static and eternal, but it evolves and consequently it has a history. But the expansion of the universe also brought with it a whole series of puzzles, the most famous one being the big bang singularity. Indeed, the equations of general relativity, together with certain assumptions about the matter content of the universe (in particular that it should obey the null energy condition, which is the assumption that the sum of energy density $\rho$ and pressure $p$ is positive) imply that the current expanding phase must be preceded by a singularity at which the spacetime curvature blows up and where general relativity predicts its own breakdown \cite{HP70}. A general expectation is that quantum effects, and in particular quantum gravity, will be able to resolve this singularity and shed light on the physics of the big bang -- a recent attempt in this direction is, for instance, provided by \cite{GT15}. However, there remains the interesting possibility that the big bang might already be resolved at the classical level, via a relaxation of the assumptions inherent in the singularity theorems. For example, one can obtain non-singular solutions in which the universe bounces instead of crunches when the null energy condition is violated \cite{BKO07,Creminelli:2007aq,Leh11}. Such solutions are of great intrinsic interest, but one may also hope that they capture salient features of quantum resolved singularities (an example of this is provided by \cite{Rovelli:2013zaa,Wilson-Ewing:2013bla}). Regardless of whether that will turn out to be the case, these solutions are appealing because they allow physical phenomena to remain fully calculable, all the way through the bounce. This is of obvious interest for cosmology, as it allows one to ask questions such as: could there have been a phase of cosmological evolution before the expanding phase (that is, before the big bang)? If so, what can we find out about this pre-expansion phase? How does it influence the post-bounce evolution?

These questions must be addressed within the context of particular models. It was long believed, for example, that violations of the null energy condition go hand in hand with the appearance of ghosts. If this were the case, the theory would be subject to a fatal growth of instabilities, and its solutions would not be trustworthy. In recent years, new matter models have been discovered, for example, the ghost condensate \cite{AHCLM04} and Galileons \cite{NRT09,NRT10,DEV09}, which in certain circumstances allow for violations of the null energy condition without the appearance of ghosts. This is already very encouraging, but nevertheless other instabilities might appear under such extreme conditions. In this context, it is important to realize that these matter models are formulated as effective theories. In order to determine their reliability it is, therefore, crucial to know their range of validity. This is the topic of the present paper -- to find out when the effective description is valid, and when it is not. This turns out to be directly related to the cosmological questions alluded to above. In particular, we want to answer the question: can a specific class of smooth,  non--singular bounces be trusted--not only at the classical level but when fluctuations in the associated scalar fields and metric are included?

In this paper, we will focus on bounces caused solely by a ghost condensate. We do this because not only do such models have the crucial property of allowing for ghost--free violations of the null energy conditions, but they are technically much simpler than pure Galileon models and mixed Galileon/ghost condensate theories. Also, ghost condensate bounces have been used in several cosmological models of interest, starting with the pre-inflationary model of Creminelli et al. \cite{CLNS06} and the new ekpyrotic cosmology of Buchbinder et al. \cite{BKO07} (see also \cite{Creminelli:2007aq}), and even have been found useful in quantum cosmology \cite{Leh15}. Furthermore they can be embedded into supersymmetry \cite{KLO11} and supergravity \cite{KLO13a,KLO14b,KLO14a}. Specifically, in the first part of \cite{KLO14a} we constructed classical bounce cosmologies based on a single real scalar field  whose kinetic terms are a ghost condensate coupled to a generalized third-order ($L_{3}$) Galileon. The scalar also possessed a potential energy of the ekpyrotic type. We analyzed the classical dynamics of this system in a flat Friedmann-Lemaitre-Robertson-Walker (FLRW) spacetime. We then went on to show that theories of this type can be generalized to $N=1$ local supersymmetry. However, in this paper we will focus solely on the non-supersymmetric theory. Furthermore, for specificity and simplicity, we will also set the coefficient of the Galileon term to zero -- that is, we will consider a scalar field with a pure ghost condensate kinetic term and an ekpyrotic potential in flat FLRW spacetime. To make this paper as self-contained as possible, we review the non-supersymmetric part of  \cite{KLO14a} in the beginning of the next section -- focussing specifically on the classical bounce solution arising from a pure ghost condensate.

Having presented this non-singular, classical bouncing cosmology, we recognize that it is essential to discuss both scalar and metric linearized perturbations in this background. This analysis is required to ensure that these perturbations do not develop large amplitudes that could disrupt the evolution of the bounce. In previous work with L.~Battarra \cite{BKLO14}, we investigated linearized perturbation theory for non--singular ghost condensate bounces where a (sub-dominant) Galileon term was also added. We demonstrated that long-wavelength co-moving curvature perturbations pass through ghost condensate bounces essentially unchanged. This remains true despite the fact that during the bounce phase the speed of sound squared $c_s^2$ becomes negative. For long-wavelength modes one can argue that the bounce occurs on a length/time scale that is so short that this cannot possibly influence the long-wavelength modes that are relevant for observations in ekpyrotic models. However, this same argument suggests that short-wavelength modes--that is, modes whose wavelengths are much shorter than the scale of the bounce--can grow significantly during the bounce phase. This leads to the first of three important questions. The first is:

\noindent $\bullet$ {\it Can the growth of these short sub-horizon co-moving curvature modes disrupt the bounce}?

The quadratic action for the co-moving curvature perturbation contains terms that are proportional to $1/H,$ where $H$ is the Hubble rate. At the bounce, the Hubble rate passes through zero and, thus, there is an apparent singularity. However, it was shown in \cite{BKLO14} that this is really only ``apparent''. An appropriate analysis reveals that the quadratic action is actually completely well-behaved and non-singular through the bounce. However, in determining the validity of the effective theory  we will have to calculate the  action to cubic order in fluctuations. This will again contain terms involving inverse powers of the Hubble rate. This leads to the second important question:

\noindent $\bullet$ {\it Will the $1/H$ terms in the cubic action just be ``apparent'' singularities, or do they signal the breakdown of the perturbative description}?

Within the context of inflation and the calculation of non-Gaussianities, the cubic action for perturbations has been calculated for a wide range of models, including ghost condensate models \cite{SL05}. The actions typically contain terms that are proportional to $1/c_s^2,$ that is, terms that are inversely proportional to the speed of sound squared. As described above, the speed of sound squared becomes negative in the vicinity of the bounce, implying that it passes through zero both before and after the bounce. Hence, there is a third important question:

\noindent $\bullet$ {\it It would appear that the cubic action becomes infinite at the moments when $c_s^2=0$, signaling the breakdown of the effective theory. Is this true -- or are these singularities only ``apparent'', disappearing upon careful calculation of the cubic action}?

We emphasize that all of the conclusions -- and questions -- just presented remain true even in the case when the coefficient of the Galileon term is set to zero. Again, to make this paper as self-contained as possible, we will review the above theory and questions in the second part of the next section -- focussing specifically on both scalar and metric linearized perturbations within the specific context of the classical bounce solution arising from a pure ghost condensate.

Having specified the results in \cite{KLO14a} and \cite{BKLO14} within the context of the pure ghost condensate theory, the bulk of this paper is devoted to examining this theory so as to answer all three of the above puzzles. We do this by calculating the strong coupling cut-off of the ghost condensate theory. We show that it can be significantly above the scale of the bounce -- so that the bounce solution can be trusted -- while still being low enough so that the dangerous short wavelength modes described above lie outside the range of validity of the effective theory. Hence, these modes can be disregarded. We also find that apparent singularities in the cubic action can be resolved by a careful calculation of the perturbative action. Our conclusion will be:  {\it there exist healthy descriptions of non-singular bounces, which can be used to replace the big bang singularity in cosmological models}. 

Finally, we note that the notation used in  \cite{KLO14a} and \cite{BKLO14} is not entirely uniform. Furthermore, some of the notation used in those papers does not conform with 
more ``standard'' notation in the cosmological literature. In order to make this paper completely consistent throughout, we use a uniform, standard notation in all of the following analysis. The relation of this notation to that used in  \cite{KLO14a} and \cite{BKLO14} should be self-evident.


\section{The cosmological model} \label{sectionmodel}

The bounce model we consider in this paper consists of a single real scalar field $\phi$ with non-canonical kinetic terms and a potential $V(\phi).$ It is identical to the model discussed in our two previous papers \cite{KLO14a,BKLO14} on bouncing cosmology, with the important exception that--for simplicity--we have set to zero the contribution from the Galileon term. In ``natural'' units -- defined by $8\pi G = M_{P}^{-2}=1$, where $M_{P}$ is the ``reduced'' Planck mass -- the Lagrangian is given by
\be
{\cal L} = \sqrt{-g}\big( \frac{R}{2} +P(X,\phi) \big), \label{Lagrangian}
\ee
where $R$ is the Ricci scalar and
\be
P(X,\phi) = \kappa(\phi) X + q(\phi) X^2 - V(\phi)
\label{AAA}
\ee
with $X \equiv -\frac{1}{2} g^{\mu\nu}{\partial_{\mu}\phi}{\partial_{\nu}}\phi.$ Since there are at most single derivatives acting on fields in the Lagrangian, it is clear that the equations of motion will be of second order. The explicit forms of the functions $\kappa, q, V$ are chosen as follows.

\begin{figure}
\centering
\subfigure[\, Kinetic functions]
  {\includegraphics[ width=0.46\textwidth]{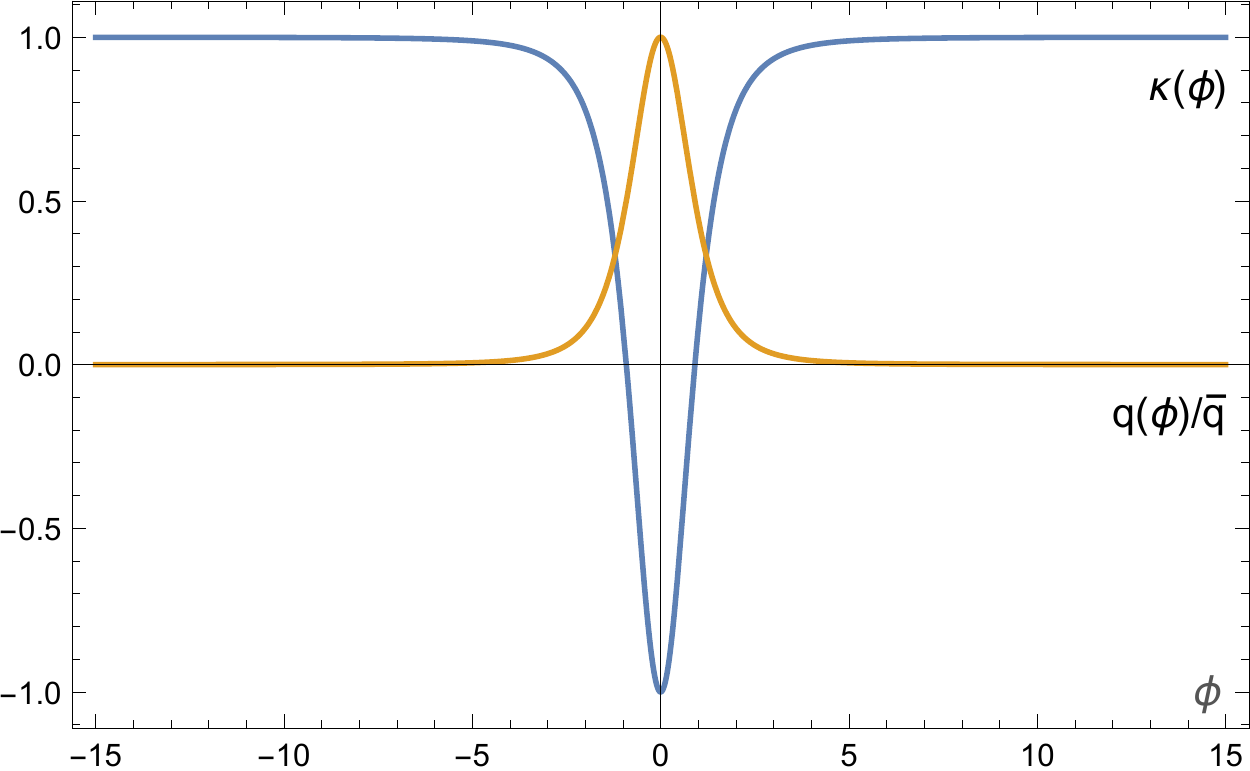} }
\subfigure[\, Ekpyrotic Potential]
 {\includegraphics[width=0.49\textwidth]{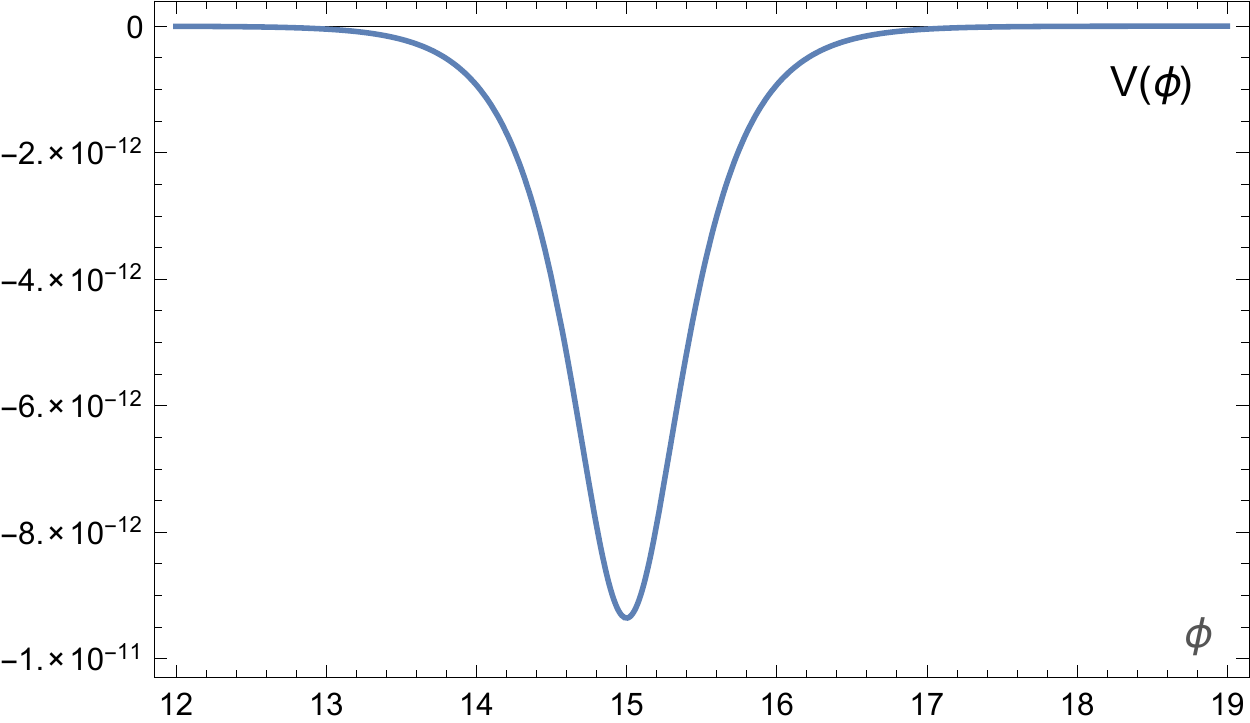}}
\caption{Graphs of the functions entering the scalar field Lagrangian. {(a)} The blue curve shows $\kappa(\phi)$ while the yellow curve shows the normalized function $q(\phi)/\bar{q}$, both with $\bar\kappa=1/4.$ {(b)} The ekpyrotic potential \eqref{Burt1} with $V_0=100, \lambda=3, \phi_{ek-end} = 15, c(\phi)=3$. The ekpyrotic phase starts at large positive $\phi,$ with the field rolling down the potential towards smaller values of the field. Around $\phi_{ek-end}$ the potential starts to come back up to zero, and is irrelevant from then on. In this model, the bounce occurs at small values, $\phi \approx 0$. }\label{fig:kaat}
\end{figure}

First, we take the kinetic function $\kappa(\phi)$ to be equal to unity everywhere except as it approaches the origin of $\phi$, where it smoothly switches sign; becoming $-1$ precisely at $\phi=0$. We use the specific form
\be
\kappa(\phi) = 1 - \frac{2}{(1+2{\bar{\kappa}} \phi^2)^2} \ ,
\label{again1}
\ee
where $\bar{\kappa}$ denotes a parameter that controls the width in field space over which the kinetic term switches sign.  This form is chosen so as to allow for a simple supersymmetric extension -- see \cite{KLO14a}. 
The function $q(\phi)$ controls the strength of the term that is the square of the ordinary kinetic term. It is chosen to interpolate between $0$ and a positive constant $\bar{q}$ in precisely the same interval where the ordinary kinetic term switches sign. We take
\be
q(\phi) = \frac{\bar{q}}{(1+2\bar{\kappa} \phi^2)^2} \ ,
\label{again2}
\ee
where, again,  we have chosen a functional form that allows for a simple supersymmetric extension.  It is crucial that this function is already non-zero when $\kappa(\phi)$ passes through zero, otherwise a singularity would develop at this point. Both functions $\kappa(\phi)$ and $q(\phi)$ are illustrated in Fig. \ref{fig:kaat}(a) for the choice
\be 
\bar{\kappa}=\frac{1}{4} 
\label{onceagain}
\ee
which, for specificity,  we will employ for the remainder of this paper. We should emphasize that the specific functions written out above are chosen for convenience of supersymmetrization only--there is, in general, considerable freedom in their choices and, in 
particular, the functional forms of $\kappa$ and $q$ need not be related in as simple a manner as they are in our example. What is important, however, is that at $\phi=0$ the kinetic part of $P(X,\phi)$ simply be
\be
P(X,0)=-X+\bar{q} X^{2} ,
\ee
that is, the canonical form for the ``ghost condensate'' \cite{AHCLM04}. It follows that in an interval containing $\phi=0$ the so-called null energy condition (NEC) is violated, thus enabling a ``bounce'' from a contracting to an expanding spacetime. 
Momentarily restoring mass dimensions in the Lagrangian density \eqref{AAA}, we see that $\bar{q}$ has mass dimension -4. It follows that the ratio of the horizon length at the bounce to the ``reduced'' Planck length is $\sim M_{P}{\bar{q}}^{1/4}$. In order for the horizon length to be sufficiently ``classical'', we want this ratio to be
\be
M_{P}{\bar{q}}^{1/4} \gtrsim 10^{2} \ ,
\label{brand1}
\ee
corresponding to a horizon mass of at most order $10^{16}$ GeV. Returning to natural units, we henceforth, for specificity, take the horizon mass to be exactly order $10^{16}$ GeV and, therefore, choose
\be
\bar{q}=10^{8} .
\label{help1}
\ee  

The potential function $V(\phi)$ is taken to be an ekpyrotic potential \cite{Khoury:2001wf} of the form
\be
V(\phi) = - V_0 v(\phi) e^{-c(\phi) \phi} ,
\label{Burt1}
\ee
where $V_{0}$ is a positive constant, $c(\phi)$ is a slowly varying function of $\phi,$ with $c(\phi) > \sqrt{6}$ over a significant field range, and
$v(\phi)$ is a function chosen so that the potential turns off for $\phi < \phi_{ek-end}$. One can take, for example, $v(\phi) = \frac{1}{2}[1+\tanh(\lambda(\phi-\phi_{ek-end}))]$ for some positive constant $\lambda$ -- see Fig. \ref{fig:kaat}(b). \\
\indent Throughout this paper, we will take the spacetime background to be a {\it flat} FLRW universe. In ``physical'' time $t$ the metric is given by
\be
\d s^2 = - \d t^2 + a(t)^2 \delta_{ij} \d x^i \d x^j \ .
\ee
We will denote derivatives with respect to the (background) physical time by $\dot{} \equiv \frac{\d}{\d t}$. The equations for the energy density, pressure and the field $\phi$ are given by
\begin{eqnarray}
3 H^2 &=& \rho = 2XP_{,X} - P \,, \\
\dot{H} &=& -\frac{1}{2} (\rho + p) = - X P_{,X} \,, \label{Hdot} \\
0 &=& P_{,\phi} - P_{,X} (\ddot\phi + 3 H \dot\phi) - P_{,XX} \ddot\phi \dot\phi^2 - P_{,X\phi} \dot\phi^2 \,.
\end{eqnarray}
These equations were analyzed in \cite{KLO14a}, where we found that at large positive values of $\phi$ the universe starts to undergo an ekpyrotic contraction phase. During this phase, the kinetic term is approximately canonical and the universe contracts slowly. For $c > \sqrt{6}$ in the potential \eqref{Burt1}, the equation of state of the scalar field satisfies $w=p/\rho > 1$  --  thus suppressing anisotropies \cite{Erickson:2003zm}.  Around $\phi=\phi_{ek-end},$ the potential bottoms out and rises back up to zero. At that time, the universe goes over into a kinetic phase; that is,  a phase where the energy density is dominated by the kinetic energy of the scalar field and the potential becomes irrelevant. Subsequently, the ordinary kinetic term switches sign while the higher-derivative term proportional to $X^2$ is switched on simultaneously. The effective ghost condensate ($P \sim -X+{\bar{q}}X^2$) leads to a brief violation of the NEC, such that the universe undergoes a ``bounce'' at small values of $\phi$ from a contracting to an expanding phase. After the bounce, the universe is in a standard expanding phase, where the kinetic term once again becomes canonical. We are assuming that 
reheating takes place around the time of the bounce, and that this causes the universe to become filled with radiation. The ordinary hot big bang cosmological model follows.\\
\begin{figure}
\centering
\subfigure[\, Scale Factor]
  {\includegraphics[width=0.50\textwidth]{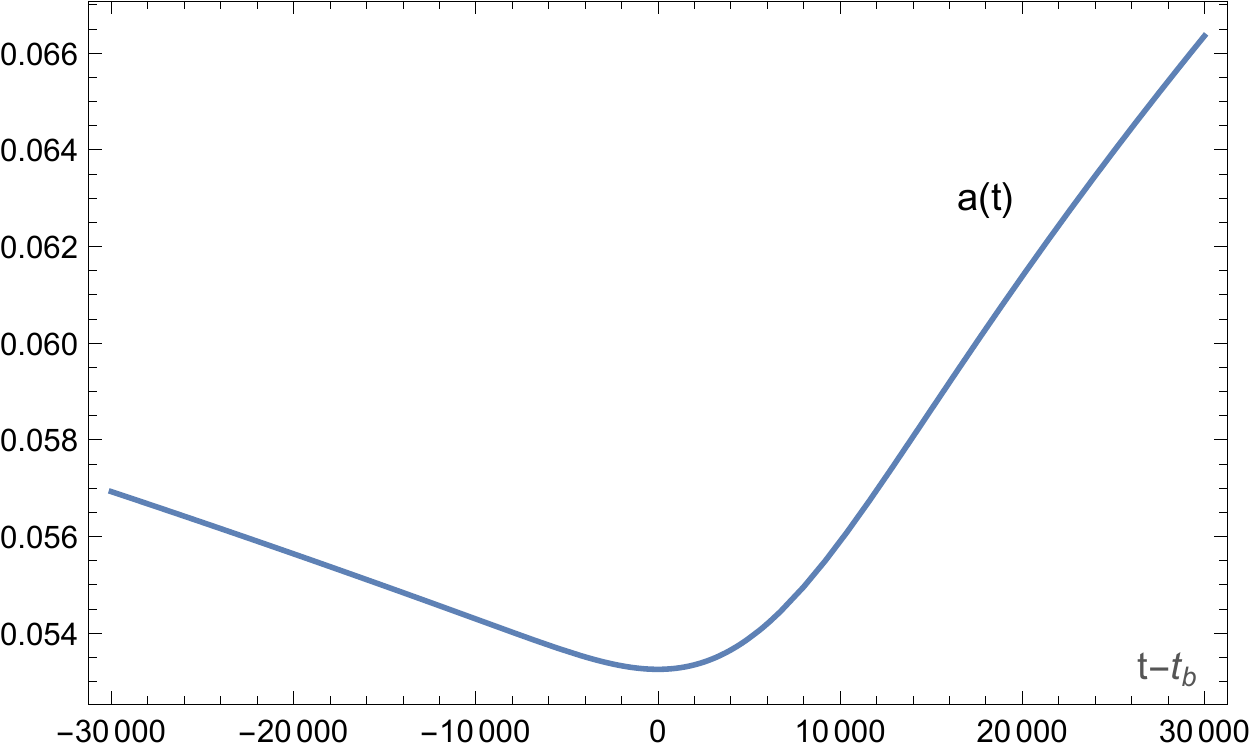}}
\subfigure[\, Scalar Field]
  {\includegraphics[width=0.49\textwidth]{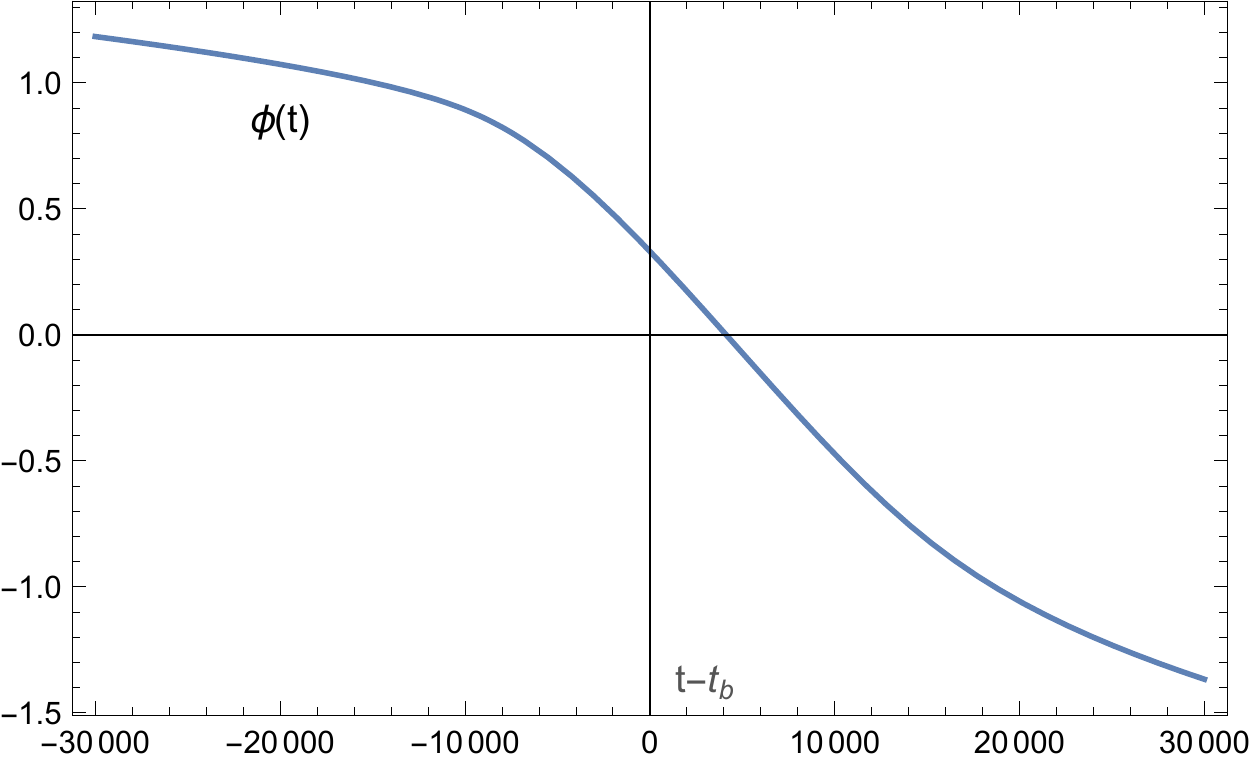}}
\caption{{(a)} The scale factor around the time of the bounce as a function of physical time $t$ minus $t_{b}$, where $t_b$ denotes the time of the bounce ($H(t_b)\equiv 0$). Our numerical evaluation starts at $\phi_0=17/2$ with $\dot\phi_0=-10^{-9},$ $a_0=1$ and $H_0$ is determined by the Friedmann equation. We are using the parameters $\bar\kappa=1/4,~\bar{q}=10^8.$ The figure shows a zoom-in on the most interesting time period, namely that of the bounce. One can clearly see that the bounce is smooth. {(b)} The evolution of the scalar field $\phi$ during the bounce phase. The approximately linear evolution near $\phi=0$ corresponds to the ghost condensate phase which is responsible for the bounce. }\label{fig:ScaleFactor}
\end{figure}
\begin{figure}
\includegraphics[width=0.55 \textwidth]{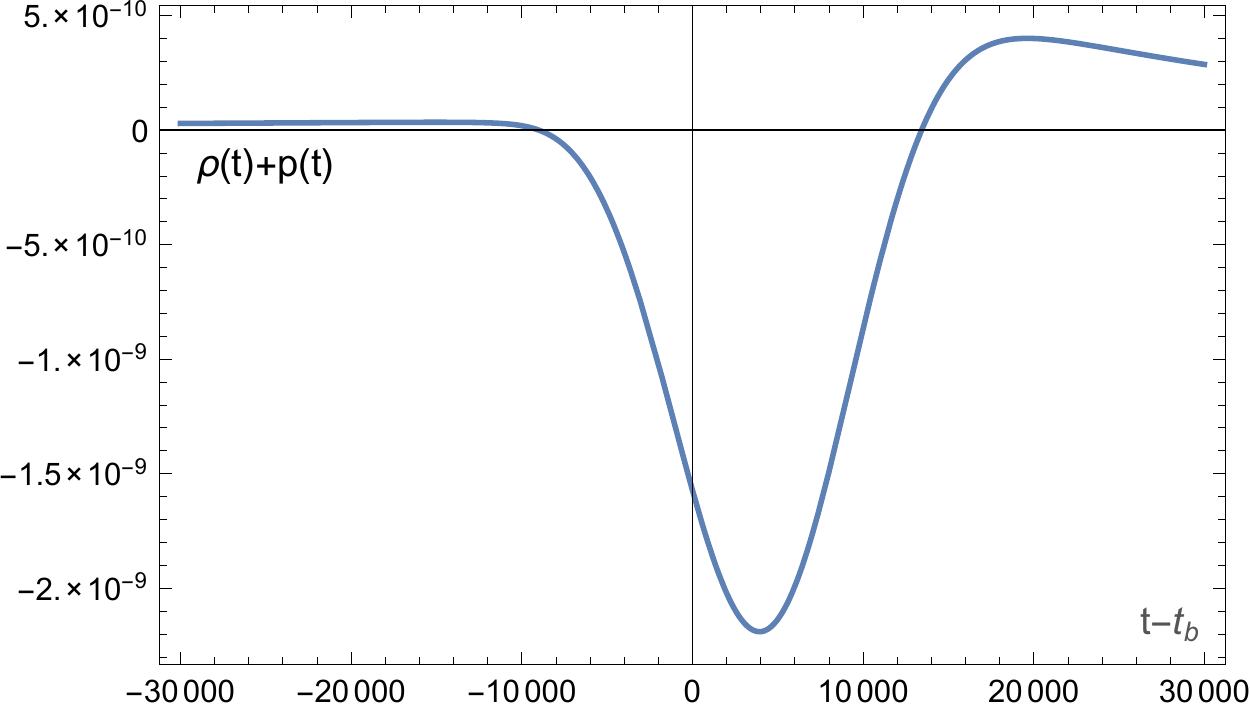}
\caption{The sum of energy density and pressure during the bounce phase. When this quantity goes negative, the null energy condition is violated. This is a necessary condition for a non-singular bounce in a flat FLRW universe, as is clear from inspecting Eq. \eqref{Hdot}.}\label{fig:NEC}
\end{figure}
\indent Figs. \ref{fig:ScaleFactor} - \ref{fig:NEC} present an explicit numerical example of the bounce phase. Here, and in the remainder of this paper, we will choose for convenience and specificity the initial conditions for our differential equations to be\footnote{These initial conditions are equivalent to those used in our earlier papers \cite{KLO14a,BKLO14}, but where $\dot\phi_0$ is re-scaled in accordance with the re-scaling of the ghost condensate mass from $\bar{q}=1$ to $\bar{q}=10^8.$ We also point out that, given that $a_0=1$, {\it the time derivative of $\phi$ takes the same numerical value in physical time, conformal time and harmonic time} at that initial moment.}
\be
\phi_{0}=\frac{17}{2}, \qquad {\dot{\phi}}_{0}=-10^{-9}, \qquad a_{0}=1 \ .
\label{BBB}
\ee
The numerical evaluation is started after the ekpyrotic phase has come to an end; that is, at the time when the kinetic phase is underway and about to go over into the bounce phase.  As the figures show, a smooth bounce is obtained during the time period that the NEC is violated. Furthermore, we note from Fig. \ref{fig:ScaleFactor}(b) that during the time that the NEC is violated, the scalar field evolves almost exactly linearly with time--this is a characteristic feature of ghost condensation. A detailed analysis in \cite{KLO14a} shows that during the bounce period, when the scalar field reaches its highest velocity, our effective field theory treatment remains consistent and applicable. We conclude that a smooth, singularity free solution of the ``classical'' field equations of Lagrangian \eqref{AAA} corresponding to a bounce from a contracting to an expanding flat FLRW spacetime exists and is trustworthy.

But what about quantum fluctuations in the scalar field and the metric? Could such fluctuations have pathologies that preclude a consistent, singularity free bouncing cosmology?
A study of the quantum perturbations of the scalar field and the scalar components of the metric in this class of bounce spacetimes was carried out in \cite{BKLO14}. Specifically, we addressed the question of the evolution of gauge invariant co-moving curvature perturbations of various wavelengths through the non-singular bounce cosmology presented above. To keep the notation in this paper consistent, we will analyze the results in \cite{BKLO14} using ``natural'' units  and the Lagrangian density given in  Eqs. \eqref{AAA}-\eqref{onceagain}. We will also choose the constant $\bar{q}=10^8$ as specified in \eqref{help1} above.

\begin{figure}
\centering
\subfigure[\, $z^2(t)$]
  {\includegraphics[width=0.49\textwidth]{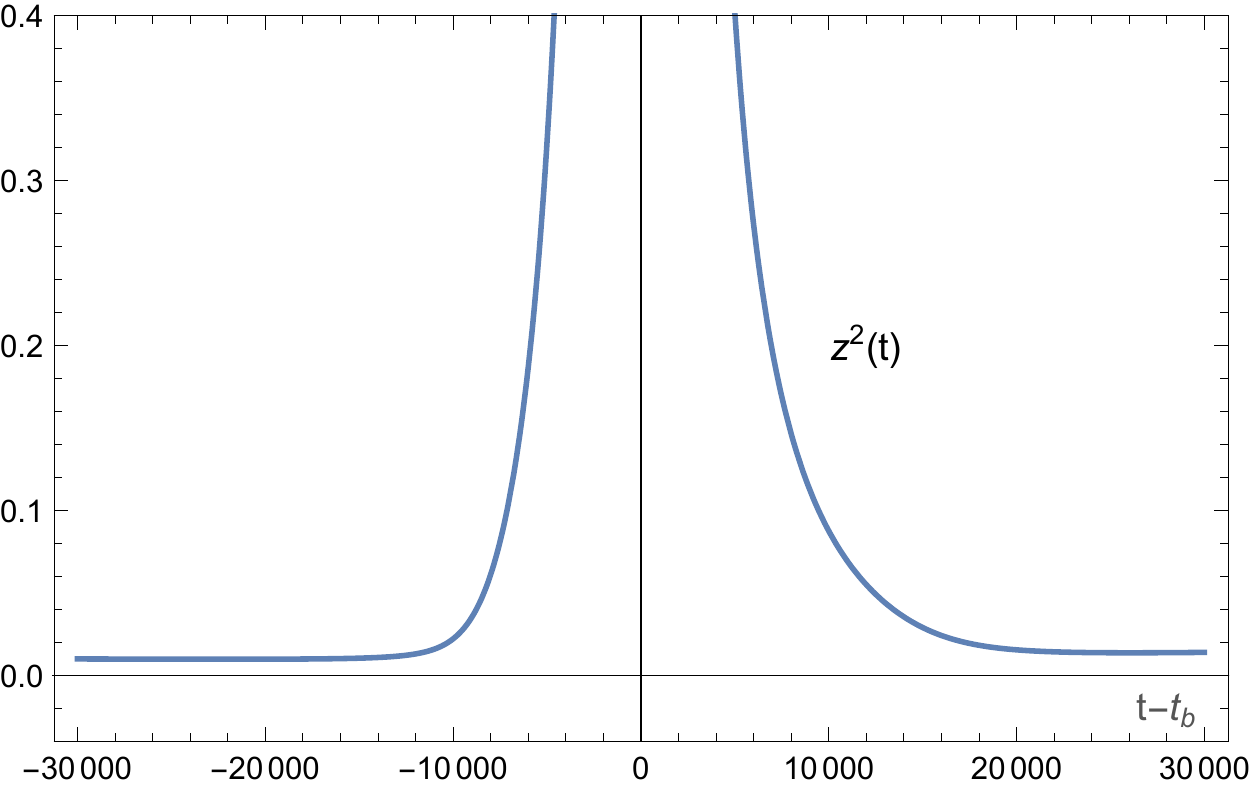}}
\subfigure[\, $c_s^2(t)$]
  {\includegraphics[width=0.50\textwidth]{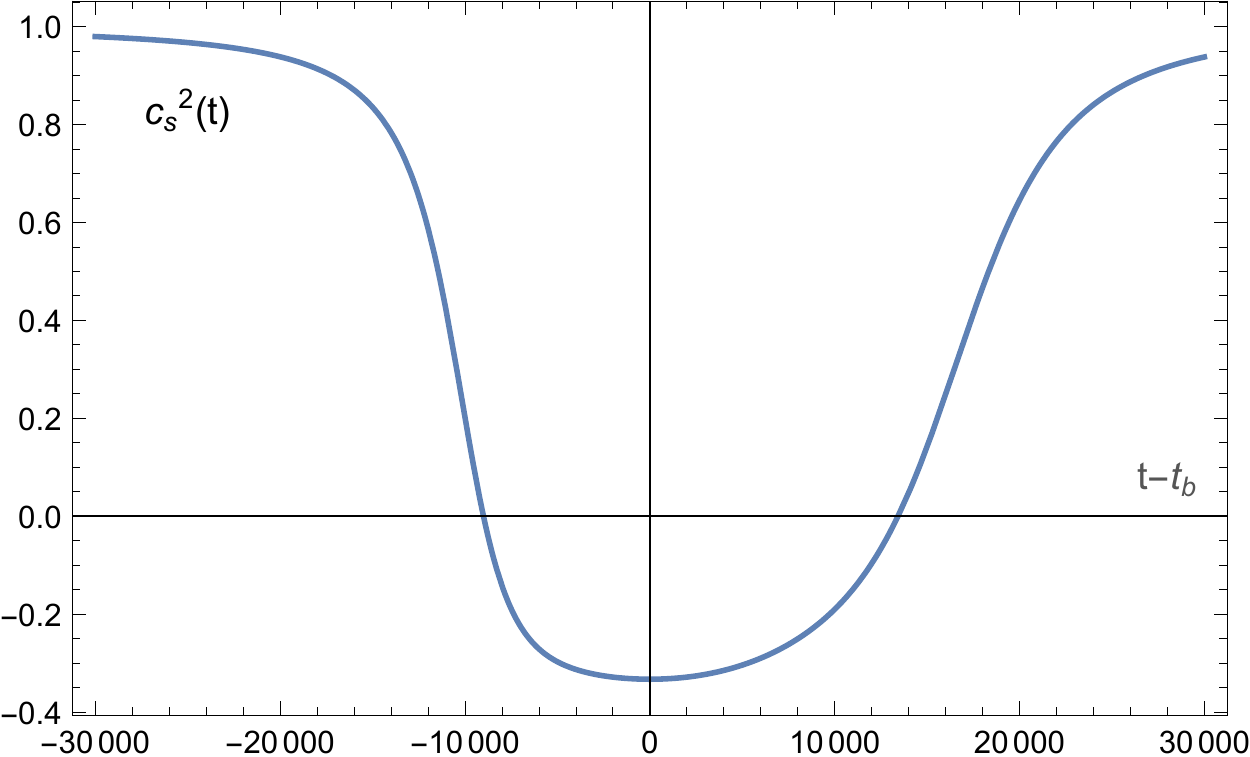}}
\caption{\label{fig:bounce3} (a) Evolution of $z ^2$ and (b) of the speed of sound squared  in the non-singular bounce background. The positivity of $z^2$ demonstrates the absence of perturbative ghost fluctuations, while the brief period over which $c_s^2$ becomes negative indicates the presence of a gradient instability.}
\end{figure}

The linearized (Fourier space) equation for the gauge invariant curvature perturbation $\cal{R}$ is given by \cite{KLO14a,BKLO14}
\begin{equation} \label{eq:conformalR}
\ddot{\mathcal{R}} + \left( 2 \frac{\dot{z}}{z} +H \right) \dot{\mathcal{R}} + c _{s} ^2 \frac{k ^2}{a^2} \mathcal{R} = 0 \ ,
\end{equation}
where $k$ denotes the co-moving wavenumber ($k/a$ thus being the physical wavenumber) and we use the definitions
\begin{eqnarray}
z ^2 & = & a^2 \frac{\Sigma}{H^2} \;,\label{eq:z2} \\
\Sigma & = &  P_{,X}X+ 2 P_{,XX} X^2 \;, \label{eq:calP} \\
c_s ^2 & = &  \frac{P_{,X} X}{ \Sigma}  \ .
\label{eq:cs2}
\end{eqnarray}
The quantities $c_s^2$ and $z^2$ are plotted in Fig. \ref{fig:bounce3}. We note that $z^2$ appears as the coefficient of the kinetic term of $\mathcal{R}$ in the perturbed action at quadratic order \cite{KLO14a} (this action will be re-derived in section \ref{sec:throughbounce}) and, thus, its positivity is essential to ensure the absence of ghosts. The plot in Fig. \ref{fig:bounce3}(a) confirms the positivity of $z^2$ and thus the absence of ghost fluctuations in this background spacetime. However, $z^2$ blows up at the bounce since the denominator of \eqref{eq:z2} passes through zero when $H=0$. Thus at the moment of the bounce the equation of motion for $\mathcal{R}$ becomes singular. This singularity turns out to be entirely harmless, but it motivated us to analyze the fluctuations in this bouncing spacetime in a manifestly non-singular manner in our earlier paper \cite{BKLO14}. Fig. \ref{fig:bounce3}(b) shows the time evolution of the speed of sound squared $c_s^2.$ During the phase where the NEC is violated, $c_s^2$ becomes negative, which is a signal of a gradient instability. Thus the last term in Eq. \eqref{eq:conformalR} switches sign, and will admit growing (as opposed to oscillatory) solutions. For long-wavelength modes (small $k$) one may argue that this effect can be ignored, but for short-wavelength modes (large $k$) one may fear that the perturbation modes become amplified to such an extent as to disrupt the background evolution. In order to circumvent the singularity of $z^2$ and to investigate the behaviour of the curvature perturbations across the bounce, we performed a calculation in harmonic gauge in \cite{BKLO14}, where the evolution of the curvature perturbations is entirely non-singular. For completeness, we repeat some of the main results here.
As just stated, it is useful to adopt harmonic gauge, in which the coordinates satisfy the defining relation
\be
\Gamma^{\mu}=g^{\rho\sigma}\Gamma_{\rho\sigma}^{\mu} \ .
\label{GGG}
\ee
For the background, this relation can be satisfied by choosing a ``harmonic'' time coordinate $t_h$ defined by
\be
\d t=a(t_h)^{3} \d t_h \ .
\label{DDD}
\ee
It follows that the flat FLRW metric becomes
\be
\d s^{2} =-a(t_h)^{6} \d t_h^{2} + a(t_h)^{2}\delta_{ij} \d x^{i} \d x^{j} 
\label{EEE}
\ee
while the associated background scalar field is
\be
\phi=\phi(t_h) \ .
\label{FFF}
\ee
The specific classical bounce solution discussed above is easily re-expressed in harmonic time. We then write the generic linearized scalar perturbations of our background fields as
\begin{eqnarray}
\d s ^2 & = & - a ^{6}(1 + 2 \mathbf{A}) \d t_h ^2 + 2 a ^{4} \mathbf{B}_{,i}\, \d t_h\, \d x ^{i} + a ^2 \Big[ \left(1- 2 \psib \right) \delta _{ij} + 2 \mathbf{E}_{, ij} \Big] \d x ^{i} \d x ^{j} \;,\\
\phi & = & \phi(t_h) + \Phib(t_h, x) \;,
\end{eqnarray}
where, for the sake of clarity, metric and scalar field perturbations are written in boldface. Furthermore, if one chooses the constraints
\begin{eqnarray} \label{eq:finalPert1burt}
0 & = & \mathbf{A}' + 3 \psib' + k ^{2} \left( \mathbf{E}' - a ^{2} \mathbf{B} \right) \ , \\  \label{eq:finalPert2burt}
0&=&\left( a ^{2} \mathbf{B} \right)' + a ^{4} \left(\mathbf{A} - \psib + k ^{2} \mathbf{E} \right) 
\end{eqnarray}
where $^{\prime} \equiv \frac{d}{dt_{h}}$, then the perturbed metric continues to satisfy condition \eqref{GGG}. This defines the ``harmonic'' gauge for the perturbation calculation.

\begin{figure}
\centering
\subfigure[\, Fourier Modes of the Curvature Perturbation]
  {\includegraphics[width=0.49\textwidth]{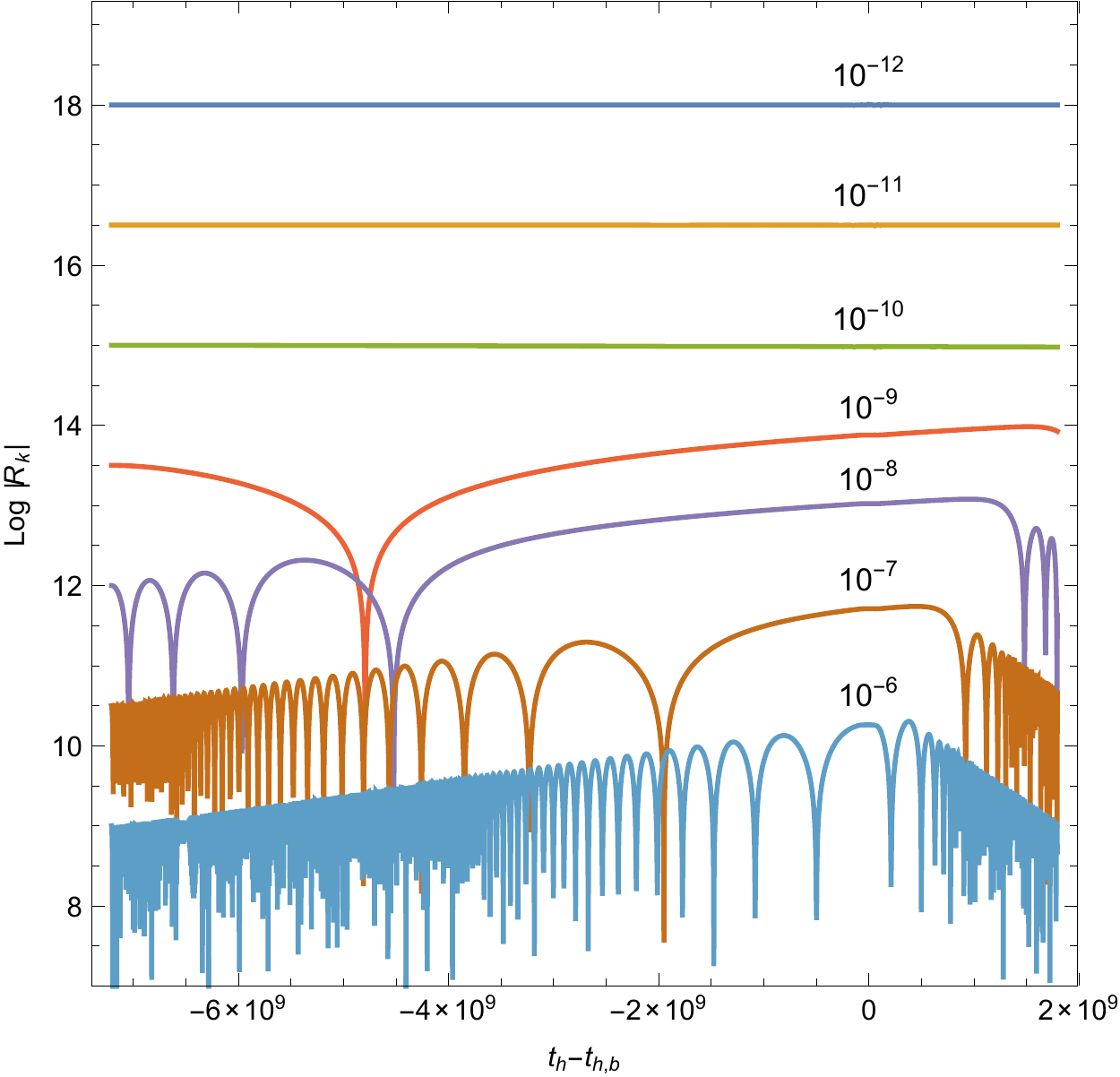}}
\subfigure[\, Physical Wavenumbers and Horizon Size]
  {\includegraphics[width=0.49\textwidth]{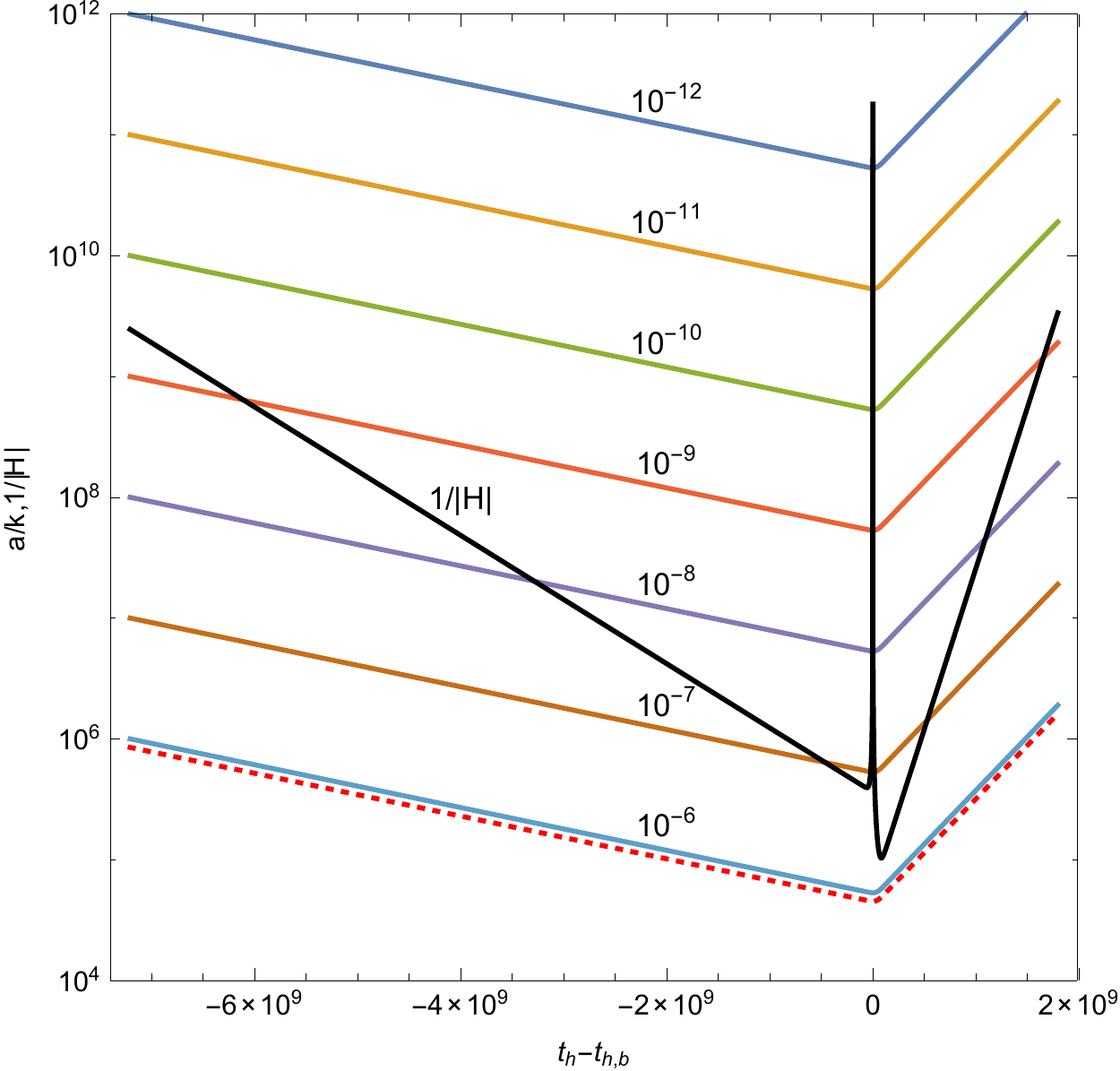}}
\caption{(a) The evolution of the co-moving curvature perturbation ${\cal R}_k$ for various co-moving wavenumbers $k$ in the range $10^{-12} - 10^{-6},$ in the bouncing background of Figs. \ref{fig:ScaleFactor} - \ref{fig:bounce3} and expressed as a function of harmonic time $t_h,$ with $t_{h,b}$ denoting the time of the bounce. This figure is adapted from \cite{BKLO14}. The initial conditions for the perturbations are chosen to correspond to the Bunch-Davies state appropriate for super-horizon perturbations, in particular ${\cal R}_k \propto k^{-3/2}.$ Long-wavelength modes evolve essentially unchanged across the bounce. (b) The horizon size $1/|H|$ (in black) vs. the various physical wavelengths $a/k$ of the perturbation modes. Modes with wavenumbers $k \leq 10^{-7}$ leave the horizon before the bounce, while shorter wavelength modes remain sub-horizon throughout. The red dotted line corresponds to a wavelength a factor of $2$ smaller than the minimum horizon size reached during the bounce phase. Its significance will become clear in section \ref{sec:SC}.}\label{fig:bounce5}
\end{figure}

The differential equations, the initial conditions and numerical solutions for the perturbation variables $ \mathbf{A}$, $ \mathbf{B}$, $\psib$, $ \mathbf{E}$ and $ \Phib$ in harmonic gauge were completely analyzed in \cite{BKLO14}. Using these results, and the definition
\begin{equation}
\mathcal{R} \equiv \psib + \frac{\cH}{ \phi'} \Phib \ ,
\end{equation}
we obtained singularity free expressions for the co-moving gauge invariant perturbations $\cal{R}$ as they enter from the contracting phase, pass through the bounce, and then exit into the expanding phase. This was accomplished for a wide range of initial parameters in the classical effective field theory--including a non-zero Galileon term. For the initial parameters being used, for specificity, in this paper -- that is, no Galileon term and
\be
\bar{\kappa}=\frac{1}{4}, \quad \bar{q}=10^{8}, \quad \phi_{0}=\frac{17}{2}, \quad {{\phi}^{\prime}}_{0}=-10^{-9}, \quad a_{0}=1
\label{HHH}
\ee
the results are plotted in Fig. \ref{fig:bounce5} for co-moving wavenumbers $k$ in the range $10^{-12} - 10^{-6},$ alongside a plot of the horizon size.

The long-wavelength modes $k=10^{-12},10^{-11},10^{-10}$ are super-horizon at all times except in the close vicinity of the bounce. Hence they can be described classically. These modes can be seen in Fig. \ref{fig:bounce5}(a) to remain constant to high precision, and show explicitly that the bounce occurs on a time-scale that is too short to affect them. This means, in particular, that the modes of interest for cosmological perturbations -- that is, modes that left the horizon about 50 to 60 e-folds earlier during the ekpyrotic phase and thus corresponding to wavenumbers $k \sim 10^{-30}$ -- pass through the bounce unchanged. This was the main finding in \cite{BKLO14} and is of crucial importance in comparing the predictions of bouncing cosmologies to observations. For $k=10^{-9},10^{-8},10^{-7}$ one can see from Fig. \ref{fig:bounce5}(b) that these modes leave the horizon only shortly before the bounce. For these wavenumbers a classical description is still fairly appropriate, and they are also little affected by the bounce.

\begin{figure}
\centering
\subfigure[\, Perturbations Modes Near Bounce]
  {\includegraphics[width=0.45\textwidth]{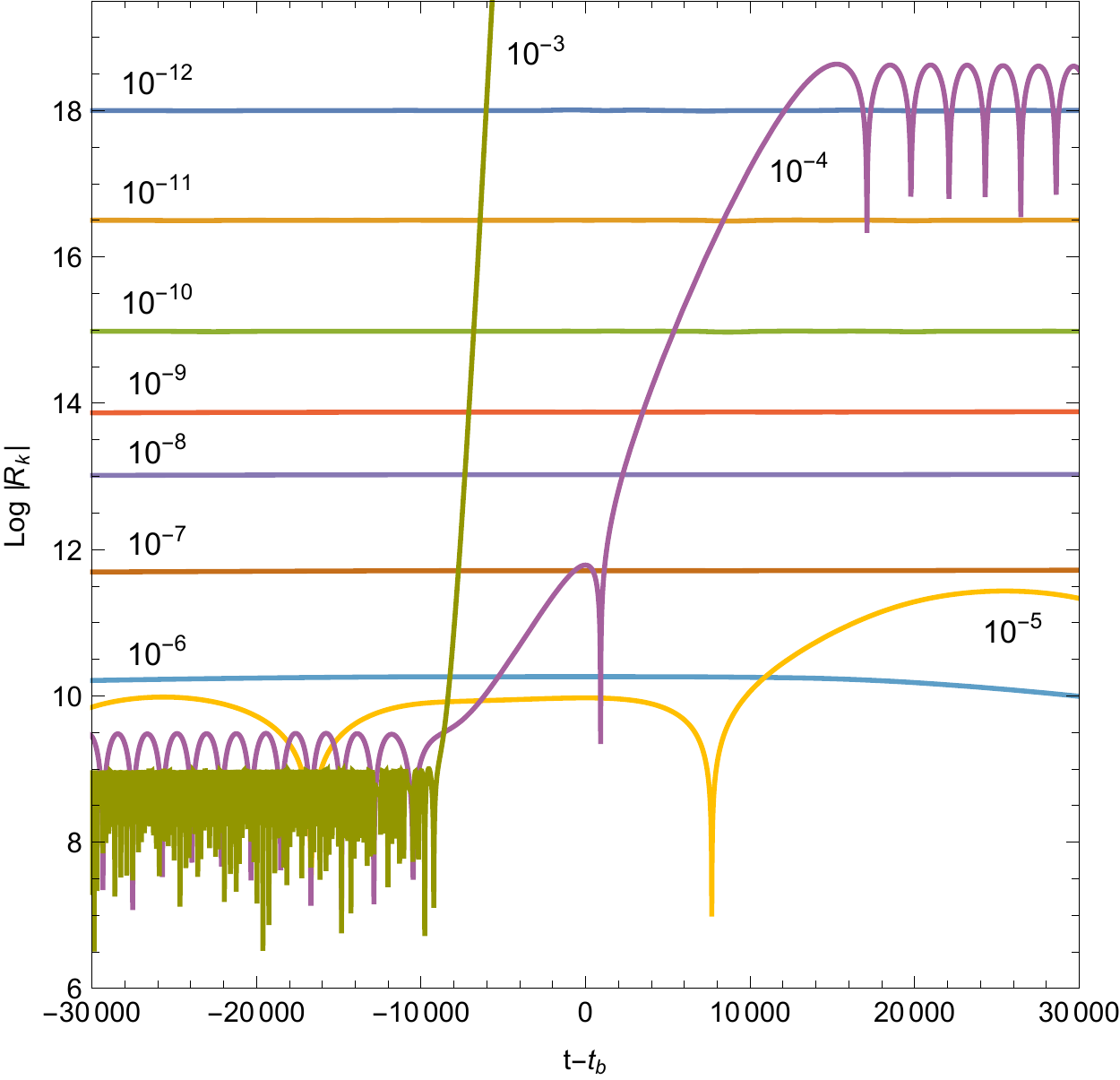}}
\subfigure[\, Amplification of Short Modes]
  {\includegraphics[width=0.45\textwidth]{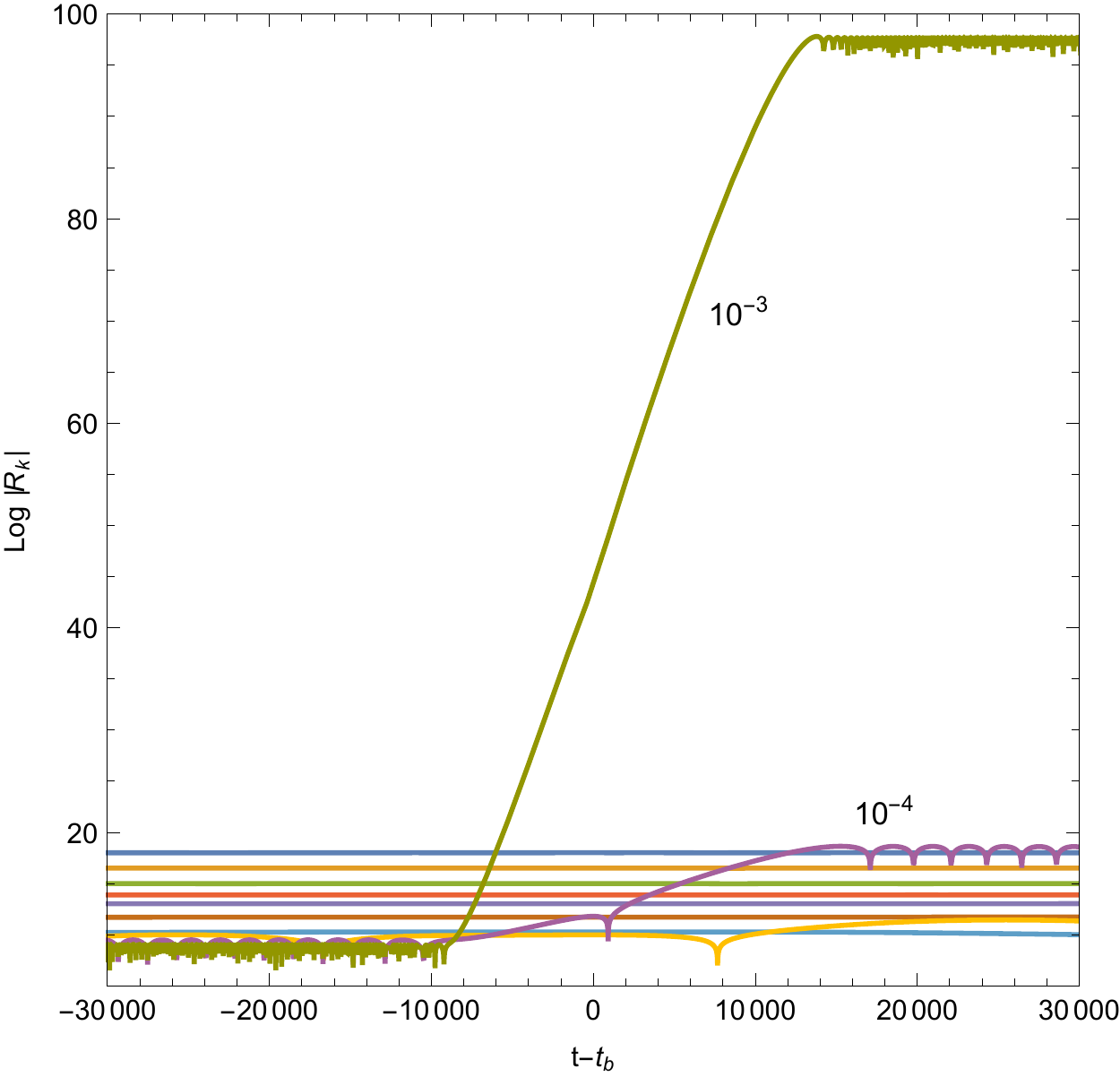}}
\caption{(a) The curvature perturbation modes ${\cal R}_k$ near the time of the bounce, expressed as functions of physical time $t.$ The initial conditions for these sub-horizon modes are taken to correspond to the early time limit of the Bunch-Davies state, in particular ${\cal R}_k \propto k^{-1/2}.$ The period of NEC violation extends from about $t=-8000$ to $t=+14000,$ as can be seen from Fig. \ref{fig:NEC}. During this time period short modes with wavenumber $k \geq 10^{-5}$ are seen to be amplified significantly. (b) The same plot, but with an expanded vertical scale. The mode with wavenumber $k=10^{-3}$ (and thus with a physical wavelength more than $3$ orders of magnitude smaller than the minimum horizon size) is seen to be amplified by almost $100$ orders of magnitude near the bounce.}\label{fig:bounce6}
\end{figure}

We strongly emphasize, however, that one cannot simply ignore the behavior of shorter-wavelength modes. For modes with $k \gtrsim10^{-6}$ the negativity of $c_s^2$ during the bounce phase becomes increasingly relevant -- see Fig. \ref{fig:bounce6} which shows examples of the behavior of short-wavelength modes during the brief time period when the NEC is violated. These short-wavelength modes remain sub-horizon into the NEC violating phase (of course, right near the bounce all modes become briefly super-horizon since $1/|H|$ momentarily blows up) and thus a classical description is inappropriate. However, as Fig. \ref{fig:bounce6} shows, these modes become increasingly amplified. For instance, the mode with $k=10^{-4}$ gets amplified by about $10$ orders of magnitude, while the mode with $k=10^{-3}$ gets amplified by nearly $100$ orders of magnitude! Such an enormous amplification makes one wonder whether these modes render a classical description of the background bouncing spacetime untrustworthy. In other words, a large amplification of short-wavelength modes may be interpreted as significant particle production -- this can potentially invalidate the bounce solution, which was obtained by solving the equations of motion in the absence of such additional matter. 

Even though the numerical solutions shown in the figures were obtained via calculations in harmonic gauge, we explicitly demonstrated in \cite{BKLO14} that the results are gauge invariant, as they should be. Thus, instead of calculating $\psib$ and $\Phib$ first (as above), we may obtain an estimate for the amplification by analyzing directly the equation of motion \eqref{eq:conformalR} for the curvature perturbation ${\cal R}$, which leads to the approximate solution
\begin{equation}
\mathcal{R}_{\text{post-bounce}} \sim \textrm{exp} \left( k \int_{ c _{s} ^2 < 0} \frac{|c_s|}{a} \d t\right) \mathcal{R}_{\text{pre-bounce}} \sim e^{ k / k_\star} \mathcal{R}_{\text{pre-bounce}}  \;.
\end{equation}
For the classical background considered here, numerical integration gives $k_\star \simeq 9\times 10^{-5}$. This equation thus gives a quasi-analytic explanation for the results shown in Fig. \ref{fig:bounce6}. More specifically, it indicates that the amplitudes for shorter wavelength modes -- that is, modes with wavelengths always smaller than the horizon (but larger than the Planck length) -- grow exponentially. Naively, this dramatic growth seems to imply that the effective field theory and, hence, the bounce solution become wildly unstable at these scales -- perhaps negating the validity of the non-singular classical bounce discussed above. It is the purpose of the present paper to prove that this is not the case and that a smooth bounce solution exists -- even including its scalar and metric perturbations.


\section{Strong coupling scale} \label{sec:SC}

The theories we are interested in are effective theories. As such, they are only valid up to some energy scale $\Lambda$ at which the fluctuations become strongly coupled. At this energy scale quantum corrections to the theory become large, and we cannot trust the tree level theory any further. Going to even higher energies would require an ultra-violet extension of the theory. However, crucially, for energies below the cut-off scale the predictions of the effective theory remain valid. One can determine the strong coupling scale by comparing the size of the coefficients of the cubic action for fluctuations to those of the quadratic action--keeping in mind that one loop corrections to scattering processes are determined by the cubic vertex. Therefore, the strong coupling scale does not just tell us where the classical description becomes hard to analyze, it also tells us the scale at which quantum corrections will strongly modify the theory itself. The physics occurring at energy scales above the strong coupling scale may be of great interest, but requires the use of a more complete theory with a higher cut-off scale. We will not attempt such an analysis in this paper.

\subsection{Lagrangian of $P(X,\phi)$ form}

As discussed above, in this paper we consider theories with a matter Lagrangian of the $P(X,\phi)$ form, where $X \equiv -\frac{1}{2} g^{\mu\nu}{\partial_{\mu}\phi}{\partial_{\nu}}\phi$ denotes the ordinary kinetic term of a scalar field $\phi$ of mass dimension $1$. This is minimally coupled to gravity, with the full action given by
\begin{equation}
S = \int \d t \d^3 x \sqrt{-g} \left( \frac{\MP^2}{2}R + P(X,\phi) \right) \,.
\end{equation} 
Note that, henceforth, we no longer use ``natural'' units but, rather, explicitly display all masses--such as the reduced Planck mass $M_{P}$. Hence, for example, the functions $\kappa(\phi)$ and $q(\phi)$ have mass dimensions 0 and -4 respectively. This will be the case for the remainder of the paper.
This class of theories includes the description of ordinary scalar fields with potentials, but also allows for ghost condensates and bounces. It is most convenient to employ the Arnowitt-Deser-Misner (ADM) decomposition of the metric,
\begin{equation}
\mathrm{d}s^2 = -N^2 \mathrm{d}t^2 + h_{ij} \left( \mathrm{d}x^i + N^i \mathrm{d} t \right) \left( \mathrm{d}x^j + N^j \mathrm{d} t \right),
\end{equation}
where $N$ represents the lapse function, $N_i$ the shift and $h_{ij}$ the metric on spatial slices of constant time. The action may then be written as
\begin{equation}
S = \frac{1}{2} \int \mathrm{d}t\mathrm{d}x^3 \sqrt{h} \left[ N \left(\MP^2 R^{(3)} + 2P(X,\phi) \right) +\frac{\MP^2}{N}\left( K^{ij}K_{ij} - K^2\right)  \right] \,,
\end{equation}
where $R^{(3)}$ is the three-dimensional Ricci scalar formed from $h_{ij}$ and where the extrinsic curvature is defined as
\begin{equation}
K_{ij} = \frac{1}{2}\dot{h}_{ij} - \frac{1}{2}N_{i,j} - \frac{1}{2}N_{j,i} + \Gamma^k_{ij} N_k\,.
\end{equation}
We are interested in determining the scale at which strong coupling occurs--that is, we are interested in determining the cut--off of the models under consideration, in order to assess the validity and reliability of particular solutions. We will focus on scalar perturbations here. In the Appendix we will treat vector and tensor perturbations, which turn out to have no influence on the bouncing solution. For the scalar perturbations, there is as always the question of which gauge to use. In our previous paper \cite{BKLO14} dealing with linearized perturbation theory, we found it convenient to work in harmonic gauge. However, in the present paper, where we need to derive the cubic action in fluctuations, harmonic gauge is too cumbersome.  We have, in fact, found it convenient to use both ``flat'' gauge (used throughout section \ref{sec:SC}) and ``co-moving'' gauge (used throughout section \ref{sec:throughbounce}), depending on which physical aspect we want to highlight. 

We will start our calculation in {\it flat gauge} where the spatial metric $h_{ij} = a(t)^2 \delta_{ij}$ is kept fixed (by choosing the appropriate time and space reparameterisations of the coordinates) as the spatial section of a flat FLRW universe. The remaining scalar perturbations are defined as
\begin{eqnarray}
\phi &=& \phi(t) + \varphi(t,x^i), \\
N &=& 1 + \alpha(t,x^i), \\
N_i &=& \partial_i \beta(t,x^i).
\end{eqnarray}
The constraints arising from varying the shift and lapse functions are
\begin{eqnarray}
&& \MP^2 R^{(3)} + 2P -4P_{,X}X - \frac{\MP^2}{N^2}\left( h^{ik} h^{jl}K_{ij}K_{kl} - K^2\right) = 0, \\ && \left[ \frac{1}{N}\left( h^{jl}K_{il} - K {\delta_i}^j\right) \right]_{\mid j} = 0,
\end{eqnarray}
where $_{\mid j}$ denotes a covariant derivative with respect to the three-dimensional metric $h_{ij}$ and $K=h^{ij} K_{ij}.$ At linear order, which is all we will need, the constraints are given by
\begin{align}
\alpha=&\ \frac{\dot\phi}{2\MP^2 H}P_{,X}\,\varphi \label{eq:alpha}\\
\frac{1}{a} \p^2\beta =&\ \left(\frac{1}{2\MP^2{H}}P_{,\phi}+\frac{\dot\phi}{2\MP^4 H^2}P P_{,X}-\frac{\dot\phi^3}{4\MP^4 H^2}P_{,X}^2-\frac{\dot\phi^2}{2\MP^2 H}P_{,X\phi}+\frac{\dot\phi^5}{4\MP^4 H^2}P_{,X} P_{,XX}\right)\varphi\notag\\
&\ +\left(-\frac{\dot\phi}{2\MP^2 H}P_{,X}-\frac{\dot\phi^3}{2\MP^2 H}P_{,XX}\right)\dot\varphi,
\end{align}
where $\partial^2 = \delta^{ij} \partial_i \partial_j$ is summed only over spatial indices and where in the constraint for $\beta$ we have already used \eqref{eq:alpha} to replace $\alpha.$
The action in flat gauge and at quadratic order in fluctuations is given by
\begin{align}\label{quad}
S^{(2)}=\int\d t\d^3x a^3&\Big\{\frac12\dot\varphi^{2}\left[P_{,X}+P_{,XX}\dot\phi^{2}\right]-\frac1{2a^2}P_{,X}(\p\varphi)^2\notag\\
&+\varphi^2\Big[\frac12 P_{,\phi\phi}+\frac{3\dot\phi^2P_{,X}^2}{8\MP^2}+\frac{\dot\phi P_{,X}P_{,\phi}}{2\MP^2 H}+\frac{\dot\phi^4 P_{,X}^3+\dot\phi^6 P_{,X}^2 P_{,XX}}{8\MP^4 H^2}+\frac{P_{,X}^2\dot\phi\ddot\phi}{2\MP^2H}\notag\\
&\phantom{+\varphi^2\big[}+\frac{PP_{,X}^2\dot\phi^2}{8\MP^4H^2}+\frac{3P_{,X}P_{,XX}\dot\phi^3\ddot\phi}{2\MP^2H}+\frac{9P_{,X}P_{,XX}\dot\phi^4}{8\MP^2}+\frac{P_{,X}P_{,XX}P\dot\phi^4}{8\MP^4H^2}\notag\\
&\phantom{+\varphi^2\big[}+\frac{P_{,XX}^2\dot\phi^5\ddot\phi}{4\MP^2H}+\frac{P_{,X\phi}P_{,XX}\dot\phi^5}{4\MP^2H}+\frac{P_{,X}P_{,XXX}\dot\phi^5\ddot\phi}{4\MP^2H}+\frac{P_{,X}P_{,XX\phi}\dot\phi^5}{4\MP^2H}\notag\\
&\phantom{+\varphi^2\big[}-\frac12P_{,X\phi}\ddot\phi-\frac12P_{,XX\phi}\dot\phi^2\ddot\phi-\frac12P_{,X\phi\phi}\dot\phi^2-\frac32P_{,X\phi}H\dot\phi\Big]\Big\}
\end{align}
The speed of propagation (speed of sound) $c_s$ of the fluctuations can be read off from the ratio of spatial to time derivative terms,
\begin{equation}
c_s^2 = \frac{P_{,X}}{P_{,X} + P_{,XX} \dot\phi^2 } \ .
\label{again3}
\end{equation}
The quadratic action shows that for an ordinary scalar field with $P=X,$ the canonically normalized perturbation variable is $\varphi.$ Note that the perturbation in the shift function ($\beta$) simply does not appear here, and the perturbation in the lapse ($\alpha$) has been eliminated via the constraint equation.

At cubic order, the action is given by
\begin{align} \label{cube}
S^{(3)}=\int\d t\d^3x a^3&\Big\{\dot\varphi^{3}\left[\frac12\dot\phi P_{,XX}+\frac1{6}\dot\phi^{3}P_{,XXX}\right]\notag\\
&+\dot\varphi^2\varphi\Big[-\frac{\dot\phi P_{,X}^2}{4\MP^2 H}-\frac{2\dot\phi^3P_{,X}P_{,XX}}{\MP^2 H}-\frac{\dot\phi^5 P_{,X}P_{,XXX}}{4\MP^2 H}+\frac12 P_{,X\phi}+\frac12\dot\phi^2 P_{,XX\phi}\Big]\notag\\
&+\dot\varphi \varphi^2\Big[\frac{\dot\phi^3P_{,X}^3}{4\MP^4 H^2}-\frac{\dot\phi^2P_{,X}P_{,X\phi}}{2\MP^2 H}+\frac12 \dot\phi P_{,X\phi\phi}+\frac{5\dot\phi^5P_{,X}^2 P_{,XX}}{8\MP^4 H^2}\notag\\
&\qquad\quad-\frac{\dot\phi^4P_{,X}P_{,XX\phi}}{2\MP^2 H}+\frac{\dot\phi^7P_{,X}^2P_{,XXX}}{8\MP^4 H^2}\Big]\notag\\
&+\varphi^3\Big[\frac16 P_{,\phi\phi\phi}+\frac{\dot\phi P_{,X}P_{,\phi\phi}}{4\MP^2 H}+\frac{3 \dot\phi^3P_{,X}^3}{8\MP^4 H}-\frac{\dot\phi^5 P_{,X}^4}{16\MP^6 H^3}+\frac{\dot\phi^4 P_{,X}^2 P_{,X\phi}}{8\MP^4 H^2}\notag\\
&\qquad-\frac{\dot\phi^3P_{,X}P_{,X\phi\phi}}{4\MP^2 H}-\frac{\dot\phi^7P_{,X}^3P_{,XX}}{8\MP^6 H^3}+\frac{\dot\phi^6P_{,X}^2P_{,XX\phi}}{8\MP^4 H^2}-\frac{\dot\phi^9P_{,X}^3P_{,XXX}}{48\MP^6 H^3}\Big]\notag\\
&+\frac{\dot\phi P_{,X}}{4 a^2 H}\varphi\big[\p^2\beta\p^2\beta-\beta_{,ij}\beta^{,ij}\big] +\varphi^2\p^2\beta\big[\frac{\dot\phi^2P_{,X}^2}{4\MP^2 a H}+\frac1{2a}\dot\phi P_{,X\phi}-\frac{P_{,X}P_{,XX}\dot\phi^4}{4\MP^2 a H}\big]\notag\\
&+\varphi(\p\varphi)^2\big[-\frac{\dot\phi P_{,X}^2}{4\MP^2 a^2 H}-\frac1{2a^2}P_{,X\phi}+\frac{P_{,X}P_{,XX}\dot\phi^3}{4\MP^2 a^2 H}\big] \notag\\
& -\dot\varphi \p\varphi\p\beta\frac{1}{a}\big[P_{,X}+\dot\phi^2P_{,XX}\big] -\frac1{2a^2}\dot\phi P_{,XX}\dot\varphi(\p\varphi)^2\Big\}
\end{align}
We are now ready to analyze various special cases of interest.

\subsection{Example of a canonical scalar field}

First, as a check on our formalism, we want to determine the strong coupling scale for a scalar field with an ordinary kinetic term plus a potential, $P(X,\phi)=X-V(\phi).$ For this case, the quadratic and cubic actions simplify to 
\begin{align}
S^{(2+3)}=\int\d t\d^3x a^3&\Big\{\frac12\left[\dot\varphi^2-\frac{1}{a^2}(\p\varphi)\right] +\varphi^2\Big[-\frac12 V_{,\phi\phi}-\frac{\dot\phi V_{,\phi}}{\MP^2 H}-\frac{V\dot\phi^2}{2\MP^4H^2}\Big]\notag\\
&-\dot\varphi^2\varphi\big(\frac{\dot\phi}{4 \MP^2 H}\big)+\dot\varphi \varphi^2\big(\frac{\dot\phi^3}{4 \MP^4 H^2}\big)\notag \\ &+\varphi^3\left(-\frac{V_{,\phi\phi\phi}}6-\frac{\dot\phi V_{,\phi\phi}}{4 \MP^2 H}+\frac{3\dot\phi^3}{8 \MP^4 H}- \frac{\dot\phi^5}{16 \MP^6 H^3}\right)\notag\\
&+\frac{\dot\phi}{4 a^2 H}\varphi\big[\p^2\beta\p^2\beta-\beta_{,ij}\beta^{,ij}\big]+\varphi^2\p^2\beta\big(\frac{\dot\phi^2}{4 \MP^2 a H}\big)\notag \\ & -\varphi(\p\varphi)^2\big(\frac{\dot\phi}{4 \MP^2 a^2H}\big)-\dot\varphi \p\varphi\p\beta \frac1a \Big\} \,,
\end{align}
while the constraint reduces to
\be
\frac1a \p^2\beta=-\frac{1}{2\MP^2 H}(V_{,\phi}+\frac{\dot\phi}{\MP^3 H}V)\varphi-\frac{\dot\phi}{2\MP^2 H} \dot\varphi \,.
\ee
The quadratic action shows that $\varphi$ is already the canonically normalized perturbation variable. One could, in principle, simplify the action further using integrations by parts. However, the main features are already clear in the present form; that is, if we define the slow--roll/fast--roll parameter 
\begin{equation}
\epsilon \equiv - \frac{\dot{H}}{H^2}\,,
\end{equation}
then we have that $\dot\phi/H = \sqrt{2\epsilon}\, \MP.$ The parameter $\epsilon$ is typically of order ${\cal O}(10^{-2}) - {\cal O}(10^2),$ where this range encompasses a free scalar, a massive scalar and typical inflationary and ekpyrotic models as well. One can see that all terms in the cubic action (including those involving $\beta$) have coefficients that are of this order or smaller (some terms are suppressed by additional factors of $\dot\phi,$ which we take to be smaller than the Planck scale in magnitude). 

The cut-off of the theory is determined by comparing the terms with the highest number of derivatives at quadratic and cubic order, since at high energies the terms with the most derivatives are the most relevant ones. Writing the dominant terms as 
\begin{align}
S^{(2+3)} & \supset \int\d t\d^3x a^3 \Big\{\frac12 \dot\varphi^2-\dot\varphi^2\varphi\big(\frac{\dot\phi}{4 \MP^2 H}\big) + \cdots \Big\} \\ & \equiv  \int\d t\d^3x a^3 \Big\{ \frac12 \dot\varphi^2 - \frac1{2\Lambda_s} \dot\varphi^2 \varphi  + \cdots \Big\}\,,
\end{align}
we can see that the strong coupling scale $\Lambda_s$ of an ordinary scalar field minimally coupled to gravity is given by
\begin{equation}
\Lambda_s = \frac{2H\MP^2}{\dot\phi} = \sqrt{\frac{2}{\epsilon}} \, \MP \,.
\end{equation}
That is, the cut-off is near the Planck scale--as intuitively expected.

\subsection{Example of a pure ghost condensate}

Another interesting example is provided by ghost condensate models, which can be used to model accelerated expansion and, with slight modifications, cosmic bounces. Let us first concentrate on the pure ghost condensate case, which allows for eternal ``self-accelerated'' solutions despite the absence of a potential. The simplest model consists in choosing the matter Lagrangian function to be $P(X,\phi)=-X + \bar{q} X^2,$ where $\bar{q}$ is a constant of mass dimension $-4.$ In a homogeneous FLRW background, the scalar equation of motion is given by
\begin{equation}
\frac{\d}{\d t}\left( a^3 P_{,X} \dot\phi \right) = 0\,.
\end{equation}
The ghost condensate solution corresponds to $P_{,X}=0$ -- that is, $X=1/(2\bar{q}).$ For this solution, the null energy condition (NEC) is marginally satisfied since the sum of energy density and pressure is zero,
\begin{equation}
 \rho + p = 2XP_{,X} = 0. 
 \label{aaa}
 \ee
 It follows that the energy density is given by 
\begin{equation}
\rho = 2XP_{,X} - P = \frac{1}{4\bar{q}}\,.
\end{equation} 
Thus $1/(4\bar{q})$ may be regarded as the energy density of the ghost condensate. If we now compare the quadratic and cubic $\dot\varphi$ terms, evaluating them on this ghost condensate background (at $P_{,X}=0$), we find the surprisingly simple result
\begin{eqnarray}
S^{(2+3)} = \int \mathrm{d} t \mathrm{d}^3x a^3 \left\{ \dot\varphi^2 
 +{ \bar{q}}^{1/2} \, \dot\varphi [\dot\varphi^2 - \frac{1}{a^2} (\partial \varphi)^2] - \frac{2}{a} \, \dot\varphi \partial \varphi \partial \beta \right\} \ ,
\end{eqnarray}
where, for definiteness we have chosen the positive sign $\dot\phi=+\sqrt{\bar{q}}$.
Almost all terms are vanishing due to the fact that $P_{,X}=0$ at ghost condensation. In particular, the coefficient of $(\partial\varphi)^2$ vanishes, indicating that the speed of sound of fluctuations is zero  around the ghost condensate\footnote{We note that one often considers the addition of higher--derivative terms $\sim (\Box\phi)^2$ which then contribute a $k^4$ term to the dispersion relation, see for example the discussions in \cite{CLNS06,BKO07}.}. The expression for the variation in the shift is also very simple,
\begin{equation}
\frac1a \partial^2 \beta = \frac{\sqrt{12}}{\MP} \, \dot\varphi \ .
\end{equation}
It demonstrates that $\partial^2 \beta$ is of the same order as $\dot\varphi.$ Given this relationship between $\beta$ and $\varphi,$ we may infer that $\dot\varphi \partial \varphi \partial \beta \sim \dot\varphi \varphi \partial^2 \beta $ in magnitude. It follows that that the term involving $\beta$ in the cubic action has a coefficient of order 1 and is, therefore, sub-dominant regarding the determination of the strong coupling scale. Taking into account that the constraint for the lapse function \eqref{eq:alpha} also implies that $\alpha \propto P_{,X} = 0$ on the ghost condensate solution, we discover an important feature: the metric perturbations decouple from the scalar field perturbations to the extent that the ghost condensate scale ${\bar{q}}^{~-1/4}$ is separated from the Planck scale $\MP$. In the cubic Lagrangian, the term involving $\dot\varphi^3$ is then the dominant term. In order to determine the strong coupling scale, we must compare its magnitude to that of the $\dot\varphi^2$ kinetic term. We obtain the canonical normalization of the fluctuation field $\varphi$ by re-scaling it to $\varphi \equiv \frac{1}{\sqrt{2}}\chi.$ It follows that we may write the dominant terms in the action at quadratic and cubic order as
\begin{align}
S^{(2+3)} & \supset \int\d t\d^3x a^3 \Big\{\frac12 \dot\chi^2+\frac12 \left(\frac{\bar{q}}{2}\right)^{1/2}\dot\chi^3 + \cdots \Big\} \\ & \equiv  \int\d t\d^3x \frac12 a^3 \Big\{ \dot\chi^2 + \frac1{\Lambda_{gc}^2} \dot\chi^3  + \cdots \Big\}\,,
\end{align}
There is an overall $\frac12 a^3$ factor multiplying the two relevant terms in the Lagrangian, which simply cancels out of their ratio.
Then the strong coupling scale $\Lambda_{gc}$  is given by
\begin{equation}
\Lambda_{gc} = \left( \frac{2}{\bar{q}} \right)^{1/4} \ .
\end{equation}
Therefore, the energy density of the background, $\rho= 1/(4\bar{q})$, and the strong coupling energy scale, $\Lambda_{gc}^4 = 2/\bar{q}$, are close together-- with the background energy density being smaller by a factor  of $8$. This order-of-magnitude difference is very important, since it allows the ghost condensate solution to (just) lie within the regime of validity of the effective theory. Below, however, we will show that this difference in energy scales can be increased through the inclusion of a potential.

\subsection{Ghost condensate bounces}

Above, we analyzed the simplest model of a ghost condensate, where the ghost condensate solution applies at all times. However, in a realistic cosmological context we are interested in the situation where the ghost condensate occurs only over a brief period of time, during which a smooth bounce may occur. This can be achieved by considering theories of the form 
\begin{equation}
P(X,\phi)= \kappa(\phi) X + q(\phi) X^2 - V(\phi)\,,
\end{equation}
where the functions $\kappa(\phi), q(\phi)$ can be chosen such that they turn the ghost condensate on and off -- such as those presented in \eqref{again1},\eqref{again2} and Fig. \ref{fig:kaat} in Section II.
In such a situation, the onset of ghost condensation is determined by the condition $P_{,X}=0.$ At that moment, it follows from \eqref{again3} 
that the speed of sound $c_s$ vanishes. Immediately afterwards, when $P_{,X}$ turns negative, the null energy condition starts being violated (since the sum of the energy density and pressure is given by $\rho + p = 2XP_{,X}$). The onset of NEC violation is the most crucial moment for at least two reasons. First, previous treatments within the context of inflation led to cubic actions containing terms proportional to $1/c_s^2,$ naively indicating a singularity when $c_s^{2}$ vanishes. We will return to this point later on. Second, during the bounce phase, the energy density of the background becomes small. This follows from the Friedmann equation  $3\MP^2 H^2 = \rho$ and the fact that $H=0$ at the bounce. Thus, any effects from the strong coupling regime are alleviated during the bounce phase.
Also, before the null energy condition is violated, we do not expect any troublesome effects, so that the most stringent constraints may be expected right at the interface between NEC preservation and the bounce phase. For these reasons, the strong coupling scale may be determined by looking at the action at quadratic and cubic order at the moment when $P_{,X}=0.$ The result is that
\begin{align}
S^{(2+3)}\mid_{P_{,X}=0}=\int\d t\d^3x a^3&\Big\{\frac12\dot\varphi^{2}\left[P_{,XX}\dot\phi^{2}\right]+\dot\varphi \varphi\big[\dot\phi P_{,X\phi}\big]+\varphi^2\big[\frac12 P_{,\phi\phi}\big]\notag \\ &+\dot\varphi^{3}\left[\frac12\dot\phi P_{,XX}+\frac1{6}\dot\phi^{3}P_{,XXX}\right]\notag\\
&+\dot\varphi^2\varphi\Big[\frac12 P_{,X\phi}+\frac12\dot\phi^2 P_{,XX\phi}\Big]+\dot\varphi \varphi^2\Big[\frac12 \dot\phi P_{,X\phi\phi}\Big]+\varphi^3\Big[\frac16 P_{,\phi\phi\phi}\Big]\notag\\
&+\varphi^2\p^2\beta\big[\frac2{a}\dot\phi P_{,X\phi}\big]-\varphi(\p\varphi)^2\big[\frac1{2a^2}P_{,X\phi}\big]-\dot\varphi \p\varphi\p\beta\frac{1}{a}\big[\dot\phi^2P_{,XX}\big]\notag\\
&-\frac1{2a^2}\dot\phi P_{,XX}\dot\varphi(\p\varphi)^2\Big\} \ ,
\end{align}
while the constraint is given by
\begin{align}
\frac{1}{a} \p^2\beta\mid_{P_{,X}=0} &= \left(\frac{1}{2\MP^2{H}}P_{,\phi}-\frac{\dot\phi^2}{2\MP^2 H}P_{,X\phi}\right)\varphi -\left(\frac{\dot\phi^3}{2\MP^2 H}P_{,XX}\right)\dot\varphi \notag \\ & = - \frac{\rho_{,\phi}}{2\MP^2 H}\varphi -\left(\frac{\dot\phi^3}{2\MP^2 H}P_{,XX}\right)\dot\varphi \,.
\end{align}
Notice that, at this moment, the coefficient of $(\partial\varphi)^2$ vanishes again, so that the speed of sound $c_s$ is zero. Nevertheless, since our formalism does not contain any $1/c_s^2$ factors, it is evident that the perturbative action remains perfectly non-singular and well-behaved. The dominant terms in the action are once again the $\dot\varphi^2$ and $\dot\varphi^3$ terms, as can be guessed from the treatment of the pure ghost condensate in the previous section. This can also be verified numerically for the bounce solutions we are interested in here. The strong coupling scale is inferred by first normalizing the quadratic action via the redefinition $\varphi \equiv (P_{,XX}\dot\phi^{2})^{-1/2} \chi,$ so that the dominant quadratic and cubic terms can be written as 
\begin{align}
S^{(2+3)} & \supset \int\d t\d^3x a^3 \Big\{\frac12 \dot\chi^2+\frac12 \frac{P_{,XX}\dot\phi+\frac13 P_{,XXX}\dot\phi^3}{(P_{,XX}\dot\phi^2)^{3/2}}\dot\chi^3 + \cdots \Big\} \\ & \equiv  \int\d t\d^3x \frac12 a^3 \Big\{ \dot\chi^2 + \frac1{\Lambda^2} \dot\chi^3  + \cdots \Big\}\,.
\end{align}
We can then  read off the strong coupling scale $\Lambda,$ with the result that
\begin{equation}
\Lambda  = \frac{(P_{,XX})^{3/4}\dot\phi}{(P_{,XX}+\frac1{3}\dot\phi^{2}P_{,XXX})^{1/2}} \approx (P_{,XX})^{1/4} \dot\phi\,.
\end{equation}
This scale should now be compared to the energy density of the background at that time, which is $\rho = -P.$ Using the condition that $P_{,X}=0$, which implies $X=-\kappa(\phi)/(2q(\phi)),$ it follows that
\begin{equation}
\Lambda^4 = \frac{2\kappa^2}{q}\,, \quad \rho = \frac{\kappa^2}{4q} + V(\phi) 
\end{equation} 
where the functions $\kappa$, $q$ and $V$ are evaluated at $\phi$ for which $P_{,X}=0$.
In the absence of a potential, we recover the same result as for the pure ghost condensate; namely that the energy density of the background is a factor of $8$ smaller than the strong coupling energy density. Thus, once again, the bounce solution just fits into the regime of validity of the effective theory. However, we now see that this (slightly uncomfortable) closeness of the two energy scales can be significantly affected by the presence of a potential. In particular, a negative potential during the bounce phase increases the separation between the energy density of the background and the strong coupling scale. The two scales can, in fact, be separated by an arbitrarily large factor --  provided the potential can approach close to the minimally allowed value of $V_{min} = -\kappa^2/(4q)$.\footnote{An even more negative potential would not allow for a bounce solution.} However, one would not want this separation to become too large either, since it is essential that the potentially dangerous ultra-short wavelength perturbation modes with large amplitude remain outside the regime of validity of the effective theory. It is interesting to note that a negative potential is natural in ekpyrotic models. Up to now it was typically assumed that this negative potential would be non-vanishing during the contracting phase--but rapidly vanish before, and be irrelevant at, the moment of the bounce \cite{Lehners:2008vx}. See, for example, Fig. \ref{fig:kaat} in Section II. Our results suggest a new perspective, in that we see here that the potential can still play an important role during the bounce phase. This has implications for ekpyrotic model building, as we will discuss in section \ref{sec:discussion} below. 

\begin{figure}
\includegraphics[width=.55\textwidth]{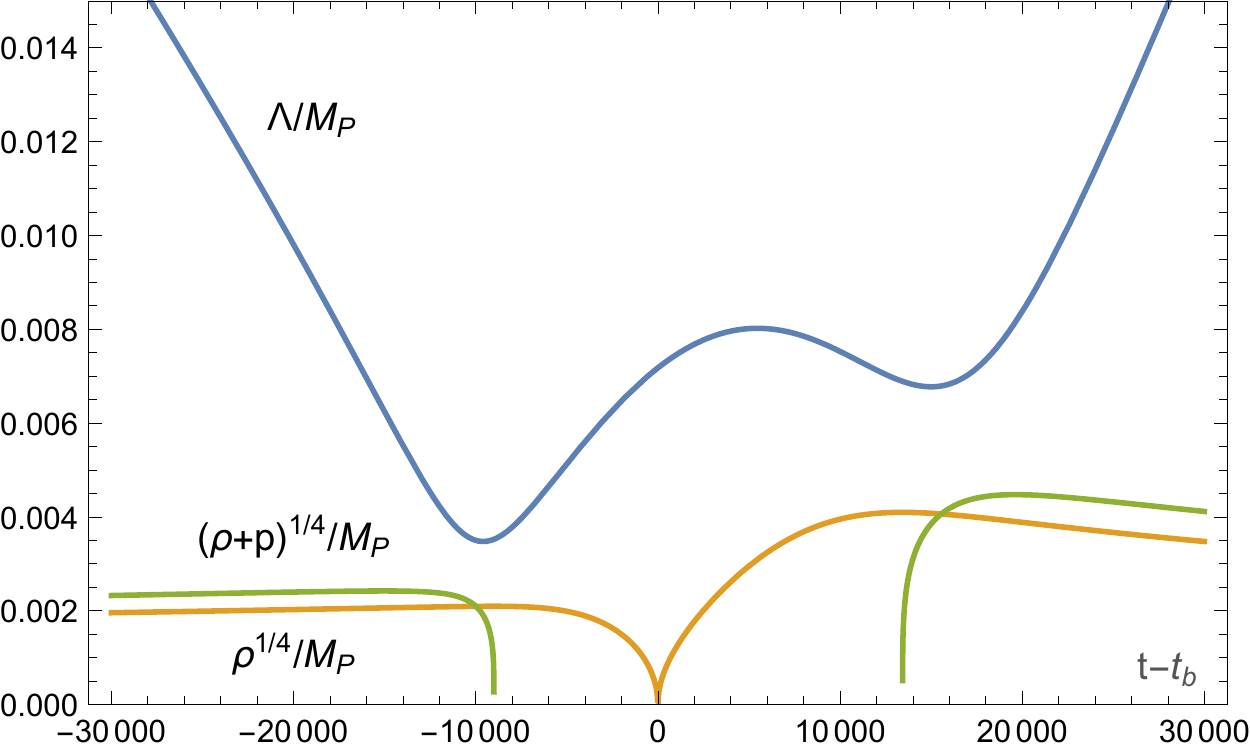}
\caption{Ghost condensate bounce without a potential, $V_0=0,$ for the bouncing background described in Section \ref{sectionmodel} and expressed in Figs. \ref{fig:ScaleFactor} - \ref{fig:bounce3}. Plotted here are the strong coupling scale $\Lambda$ and the energy density $\rho^{1/4}$ against physical time $t,$ relative to the time of the bounce $t_b.$  Also plotted is the sum of the energy density and pressure (to the quarter power). At the two moments where this quantity vanishes the null energy condition is marginally satisfied, while in the time interval in between the NEC is violated. This plot confirms that $\Lambda$ and $\rho^{1/4}$ are closest to each other precisely at the moments when the NEC starts and ends being violated. }
 \label{FigSBnopot}
\end{figure}
\begin{figure}
\includegraphics[width=.55\textwidth]{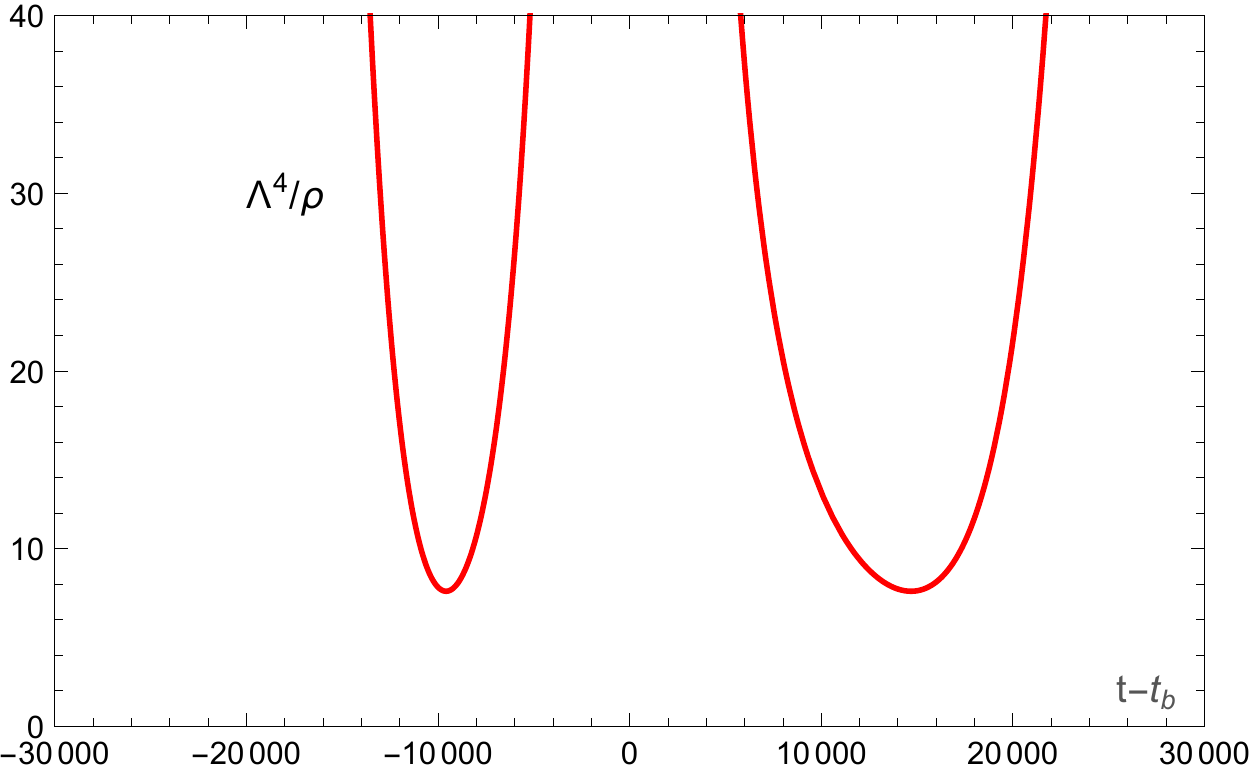}
\caption{Plot of the ratio of strong coupling scale to background energy density (to the quarter power) against physical time. As expected from our analytical treatment, we see that at the moments where the NEC starts and ends being violated, this ratio reduces to a factor of $8.$ Thus the bouncing background solution lies within the regime of validity of the effective theory, while dangerous short wavelength modes lie outside.} \label{FigSBsafetynopot}
\end{figure}

One should verify that the most stringent constraint indeed arises at the moment where $c_s^2 =0.$ We do this by numerically evaluating the strong coupling scale for a time period starting before, passing through, and then ending after the interval of NEC violation. From the $\dot\varphi^2$ and $\dot\varphi^3$ terms in the general actions \eqref{quad} and \eqref{cube}, we find -- after normalizing the scalar field as above -- that the strong coupling scale is given by
\begin{equation} \label{SCgeneral}
\Lambda = \frac{(P_{,X}+P_{,XX}\dot\phi^{2})^{3/4}}{(\dot\phi P_{,XX}+\frac1{3}\dot\phi^{3}P_{,XXX})^{1/2}} \ .
\end{equation}
Note that we have now reinstated the $P_{,X}$ term. This was set to zero above where we limited the calculation precisely to the times when $P_{,X}=0$. We can now check numerically that the energy density of the background solution comes closest to the expression \eqref{SCgeneral} precisely when $P_{,X}$ passes through zero. The specific example was introduced in the beginning of Section II. That is, we will choose the super-bounce model \cite{KLO14a}, but with the Galileon term set to zero and the coefficient $\bar{\kappa}=1/4$. Additionally, we take the coefficient $\bar{q}=10^{8}~M_{P}^{-4}$. It then follows that the kinetic part of the Lagrangian is specified by 
\begin{eqnarray}
P(X,\phi) &=& \kappa(\phi) X +q(\phi) X^2 - V(\phi) \label{eq:model1}  \ ,\\
\kappa(\phi) &=& 1- \frac{2}{(1+\frac{1}{2}\frac{\phi^2}{M_{P}^{2}})^{2}}  \ ,\\
q(\phi) &=& \frac{10^{8}M_{P}^{-4}}{(1+\frac{1}{2}\frac{\phi^{2}}{M_{P}^{2}})^2} \ .\label{eq:pot}
\end{eqnarray}
The potential energy is chosen to be in the generic form presented in \eqref{Burt1}. However, for reasons to become clear, here we take the associated functions to be $c=\sqrt{20}$ (which satisfies the ekpyrotic constraint that $c > \sqrt{6}$) and $v(\phi)=2/(1+e^{-2\sqrt{20} \frac{\phi}{\MP}})$. It follows that the potential energy can be expressed as
\be
V(\phi) = - \frac{2V_0}{e^{-\sqrt{20}\frac{\phi}{M_{P}}}+ e^{\sqrt{20}\frac{\phi}{M_{P}}}} \ ,
\label{lamp1}
\ee
where $V_{0}$ has mass dimension 4. Eq. \eqref{lamp1} is of a form previously used by Cai et al.~\cite{CEB12} in their closely related bounce model. Note that the kinetic function $\kappa$ switches sign, thereby allowing the null energy condition to be violated and, thus, enabling the presence of bouncing solutions. That we do not end up with ghost fluctuations is due to the second kinetic function $q(\phi)$, which contributes fluctuations of sufficiently large positive energy during the bounce phase.  

We first consider the example of a ghost condensate induced bounce without a potential; that is, $V_0=0$. The corresponding plots for the strong coupling scale \eqref{SCgeneral} and the energy density of the background are shown in Fig.~\ref{FigSBnopot}. Moreover, the ratio between the two scales is plotted in Fig. \ref{FigSBsafetynopot}. The plots clearly show that the most stringent moments are indeed those where the NEC starts and ends being violated. Moreover, the strong coupling scale $\Lambda^4$ is larger than the background energy density $\rho$ by a factor of $8$ precisely at those moments, as expected. 

\begin{figure}
\includegraphics[width=.55\textwidth]{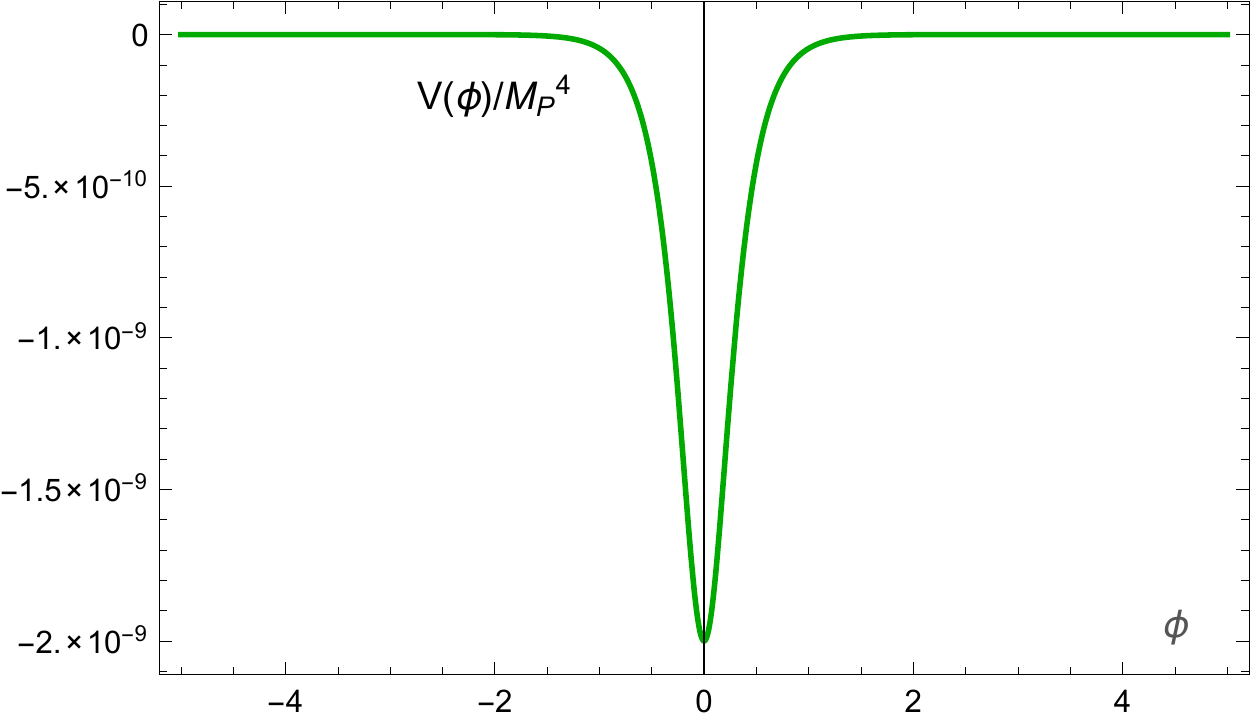}
\caption{Plot of the ekpyrotic-type potential \eqref{eq:pot} with $V_0=0.2\times 10^{-8}~\MP^4$.} \label{FigSBPot}
\end{figure}

\begin{figure}
\centering
\subfigure[\, Scale Factor]
  {\includegraphics[width=0.49\textwidth]{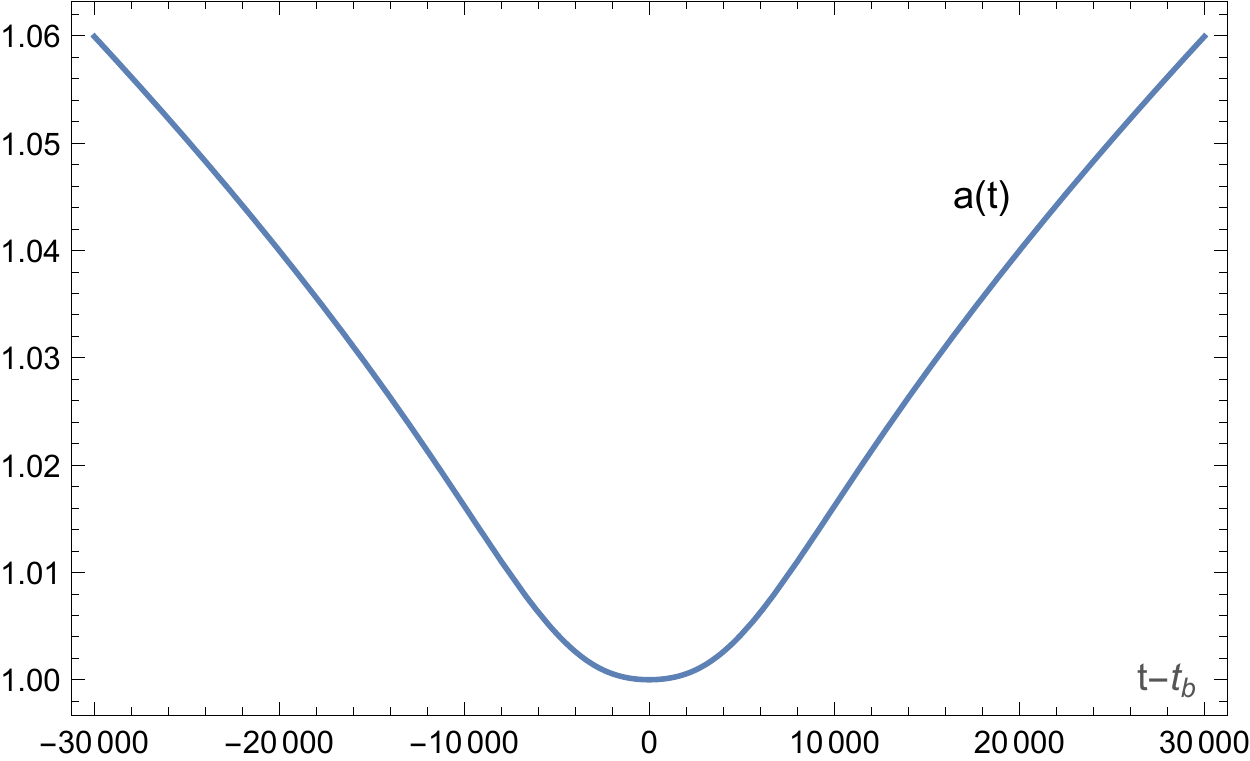}}
\subfigure[\, Scalar Field]
  {\includegraphics[width=0.49\textwidth]{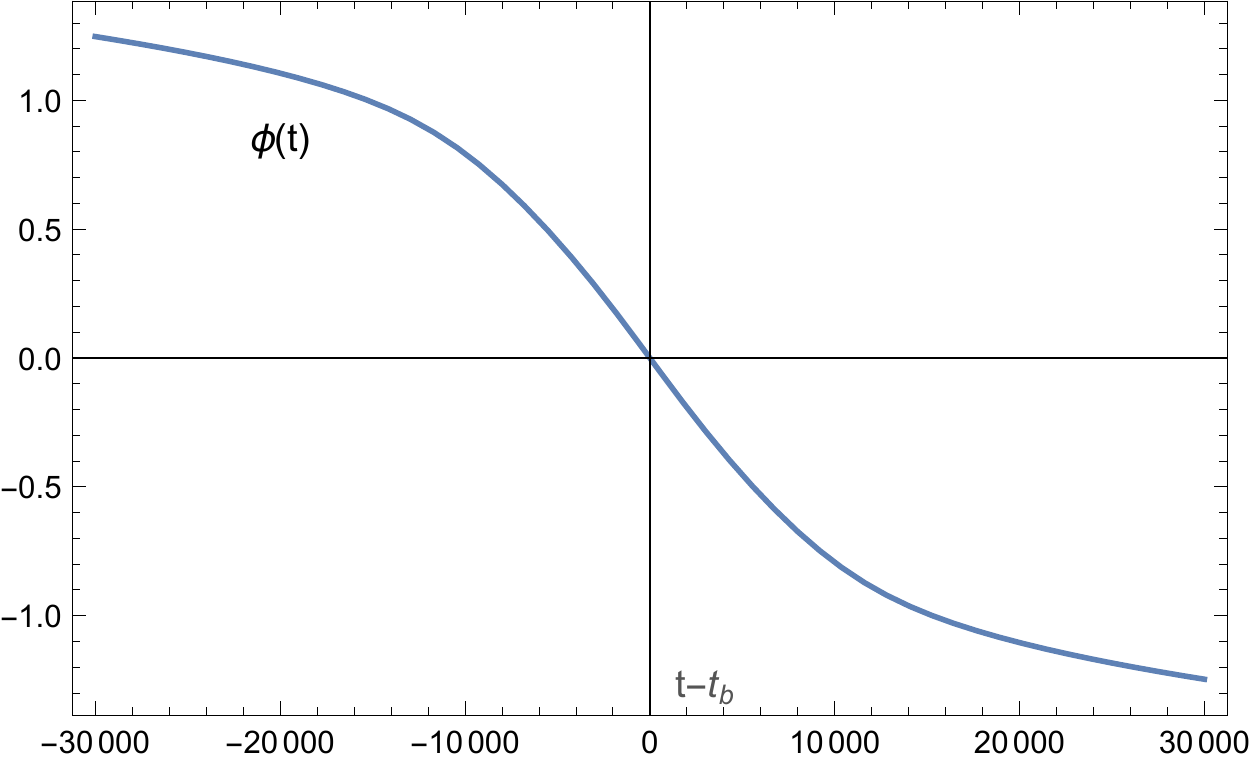}}
\caption{The bouncing background solution with the new potential Eq. \eqref{eq:pot} and $V_0=0.2~\times~10^{-8}~\MP^4$. We have fixed the initial conditions at the moment of the bounce, $H=0,$ and have chosen $\phi_{bounce}=0$. The Friedmann equation then determines the time derivative of the scalar field at that moment, since it implies $0 = 3\MP^2 H^2\mid_{bounce} = \rho\mid_{bounce} = -\frac12 \dot\phi_{bounce}^2 + \frac34 \dot\phi_{bounce}^4 - V_0.$ Fig. \ref{fig:new} (a) shows the scale factor around the time $t_{b}$ of the bounce as a function of $t-t_{b}$. Fig. \ref{fig:new} (b) shows the evolution of the scalar field $\phi$ during the bounce phase.}\label{fig:new}
\end{figure}

We now analyze how these results are modified when a non-vanishing potential is added. The potential \eqref{lamp1} we have chosen is of the ekpyrotic form, and turns on and then off symmetrically around the bounce. For specificity, we will choose $V_{0}=0.2 \times 10^{-8}~M_{P}^{4}$. This potential is plotted in Fig. \ref{FigSBPot}. The numerically evaluated background solution for the scale factor and the scalar field is displayed in Figs. \ref{fig:new}(a) and (b) respectively. These plots are qualitatively very similar to the case in Fig. \ref{fig:ScaleFactor} where the potential is absent. The strong coupling scale and background energy density for $V_0=0.2\times 10^{-8}~M_{P}^{4}$ are shown in Fig. \ref{FigSBpot}, while the ratio between these two scales is plotted in Fig.~\ref{FigSBsafetypot}. As can be seen, the strong coupling scale is now further separated from the background energy density. In this specific example, the ratio $\Lambda^{4}/\rho$ is always bigger than a factor of about $40$. This corresponds to a factor of about $2.5$ in frequency. Thus, perturbation modes with a wavelength at least $2.5$ times smaller than the horizon size at the onset of the bounce are beyond the cut-off of the theory. It follows that the modes whose amplitudes grow dangerously during the bounce period--that is, modes with wavelengths more than two orders of magnitude shorter than the horizon size at the onset of the bounce--are well outside of the range of validity of the effective theory. Hence, the bouncing spacetime solution can be trusted. Note that an even more negative potential would enhance the separation between the two scales further. As long as this separation remains smaller than a factor of about two orders of magnitude in frequency, that is, a factor $10^8$ in energy density, one need not worry about potentially dangerous short wavelength modes. For such theories the bouncing spacetime solution remains trustworthy. 

The results of this Section definitively answer the first of the three important questions that were discussed in the Introduction. That is

\noindent $\bullet$ {\it Can the growth of these short sub-horizon co-moving curvature modes disrupt the bounce}?

\noindent The answer is no -- the short wavelength sub-horizon co-moving curvature modes with amplitudes sufficiently large to disrupt the bouncing cosmology all lie in the region of strong coupling, where the effective action is no longer valid. One may now go back to Figs. \ref{fig:bounce5} and \ref{fig:bounce6} to see how this result affects the interpretation of the graphs shown there. In particular, the previous discussion has led to the conclusion that the strong coupling scale is about a factor of $2$ smaller in size than the minimum horizon size reached during the bounce phase. This scale is plotted via the red dotted line in Fig. \ref{fig:bounce5}(b). Perturbations modes with longer wavelengths ($k \leq 10^{-6}$ in that example) form a part of the effective theory, but are little affected by the bounce, while modes with shorter wavelengths ($k > 10^{-6}$) lie outside of the range of validity of the effective theory, and thus their dramatic growth can be ignored.

\begin{figure}
\includegraphics[width=.55\textwidth]{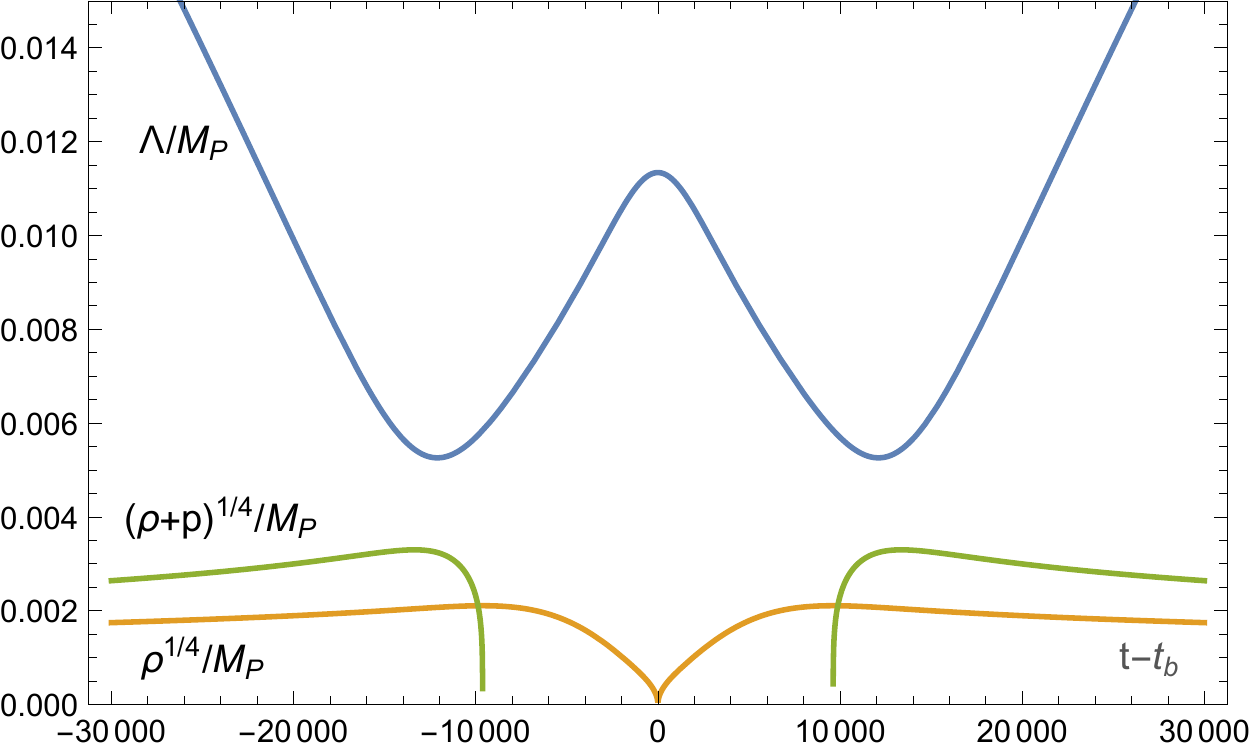}
\caption{Analogous plot to Fig. \ref{FigSBnopot}, but with a potential of strength $V_0=0.2\times 10^{-8} ~M_{P}^{4}$ included.  Again the zeroes of the curve plotting $(\rho+p)^{1/4}$ indicate the start and end of the NEC violating phase.} \label{FigSBpot}
\end{figure}
\begin{figure}
\includegraphics[width=.55\textwidth]{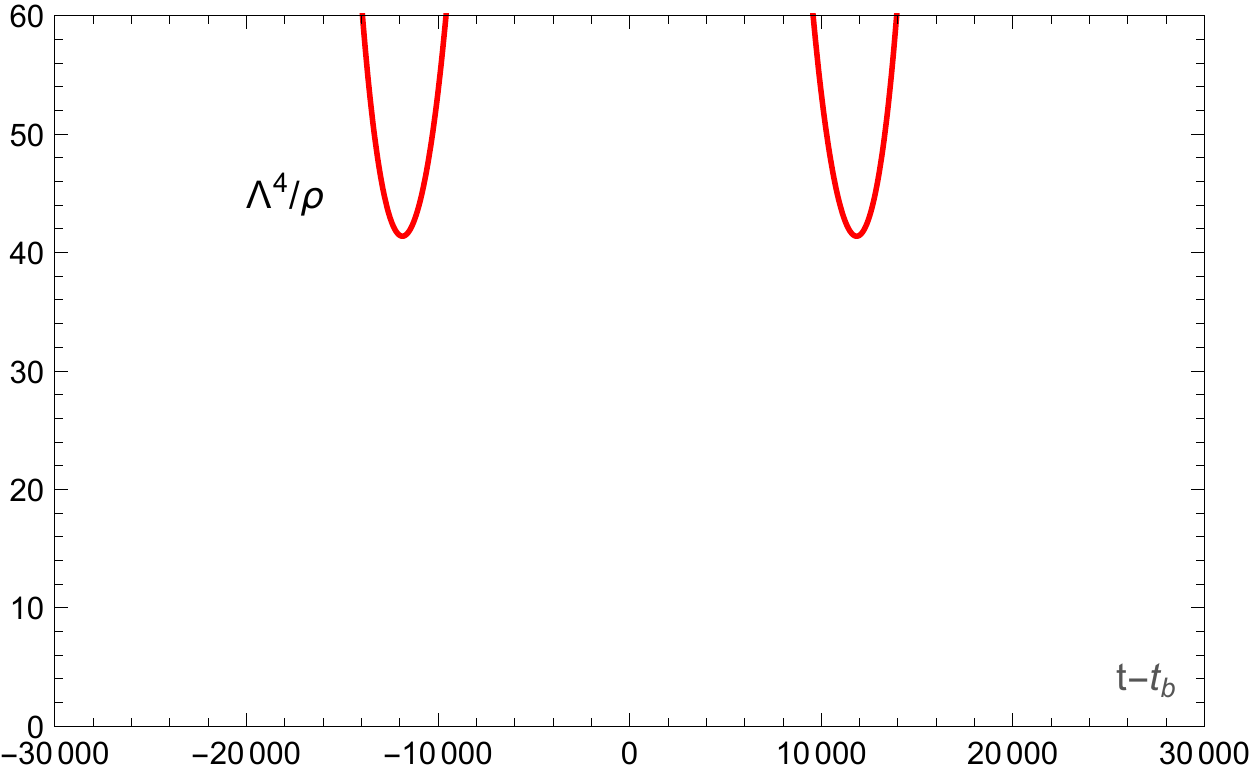}
\caption{When a negative potential is included, the background energy density and the strong coupling scale are further separated from each other. For $V_0=0.2\times 10^{-8}~M_{P}^{4}$ the ratio $\Lambda/\rho^{1/4}$ always remains above a factor of about $40.$ This implies that the background solution lies more comfortably inside the regime of validity of the effective theory, compared to the case where no potential is present during the bounce.} \label{FigSBsafetypot}
\end{figure}


\section{Through the bounce} \label{sec:throughbounce}

We now want to address the remaining two questions presented in the Introduction regarding ``apparent'' singularities in the perturbative action. In the above analysis--which was carried out in flat gauge--the action contained terms inversely proportional to the Hubble rate. Thus, it appears that a perturbative treatment might break down in the vicinity of the bounce, exactly the period we are most interested in. It turns out that it is rather difficult to prove directly in flat gauge that these are simply apparent singularities. It is, in fact, much easier to prove this by calculating the action for the co-moving curvature perturbation in co-moving gauge. Note that, as shown by Maldacena \cite{Mal03}, it is possible to transform the perturbative action from flat gauge to co-moving gauge via a time reparameterization. 
However, the re-writing of the perturbative action is highly non-trivial, as it involves many integrations by parts and the use of the perturbative equations of motion. It is, in fact, much easier to directly calculate the cubic action in co-moving gauge, which is what we now do.

In {\it co-moving gauge} the scalar field perturbation is set to zero, 
\begin{equation}
\delta \phi = 0 \ , 
\end{equation} 
so that hypersurfaces of constant scalar field are also hypersurfaces of constant time.
Again employing the ADM formalism, one can implement co-moving gauge by expanding the lapse, shift and spatial metric as
\begin{eqnarray}
N &=& 1 + \alpha(t,x^i), \\
N_i &=& \partial_i \beta(t,x^i), \\
h_{ij} &=& \delta_{ij} a^2(t) e^{2{\cal R}(t,x^i)},
\end{eqnarray}
where ${\cal R}$ is the co-moving curvature perturbation. The extrinsic curvature is then given by
\begin{equation}
K_{ij} = h_{ij} \left( H + \dot{\cal R}\right) - \beta_{,ij} + {\cal R}_{,i} \beta_{,j} + {\cal R}_{,j} \beta_{,i} - \delta_{ij} \partial {\cal R} \cdot \partial \beta\,,
\end{equation}
where $ \partial {\cal R} \cdot \partial \beta \equiv \delta^{ij} \partial_i {\cal R} \partial_j \beta$.\footnote{ We will sometimes write summed spatial indices on the same line, where it is understood that they are contracted with the Kronecker delta alone.} At linear order, the constraints are given by \cite{SL05}
\begin{eqnarray}
&& -\MP^2 \partial^2 {\cal R} - \MP^2 H \partial^2 \beta + a^2 \left[ 3\MP^2 H \left( \dot{\cal R} - H \alpha \right) + \left( P_{,X} X + 2 P_{,XX} X^2 \right) \alpha \right] = 0, \label{LinearConstraintN} \\ && \alpha = \frac{\dot{\cal R}}{H}\,. \label{LinearConstraintNi}
\end{eqnarray}
At quadratic order in fluctuations, plugging in the constraints and discarding total derivatives, the action becomes
\begin{eqnarray}
S^{(2)} &=& \int \mathrm{d}t \mathrm{d}x^3 a^3  \left[ \frac{P_{,X}X + 2P_{,XX}X^2}{H^2} \left( \dot{\cal R}\right)^2 + \frac{\MP^2 \dot{H}}{a^2 H^2}(\partial {\cal R})^2  \right] \\ &=&  \int \mathrm{d}t \mathrm{d}x^3 a^3  \left[ (P_{,X}X + 2P_{,XX}X^2) \left( \frac{\dot{\cal R}}{H}\right)^2 + \frac{2\MP^2}{a^2}\partial\left(\frac{\dot{\cal R}}{H}\right)\cdot \partial {\cal R}  + \frac{\MP^2}{a^2} \left( \partial {\cal R}\right)^2\right] \ . \label{QuadraticRewritten}
\end{eqnarray} 
The reason for rewriting the last term using integration by parts will become obvious momentarily. Often the quadratic action is expressed as
\begin{equation}
S^{(2)} = \int \mathrm{d}t \mathrm{d}x^3 a^3 \frac{\Sigma}{H^2} \left[ \left( \dot{\cal R}\right)^2 - \frac{c_s^2}{a^2} (\partial {\cal R})^2 \right] \ ,
\end{equation}
where we employ the conventional definitions 
\be
\Sigma \equiv P_{,X}X + 2P_{,XX}X^2 , \quad  c_s^2 \equiv P_{,X}/(P_{,X} + 2P_{,XX}X) = - \MP^2 \dot{H}/\Sigma \ .
\label{lamp3}
\ee
The linearized equation of motion for the curvature perturbation is then given by
\begin{equation}
H \frac{\mathrm{d}}{\mathrm{d}t}\left( \frac{\dot{\cal R}}{H} \right) + \left( 3H^2 + H \frac{\dot\Sigma}{\Sigma} - \dot{H}\right) \frac{\dot{\cal R}}{H} + \frac{\MP^2 \dot{H}}{a^2 \Sigma}\partial^2 {\cal R} = 0\,.
\end{equation}
At the bounce, where $H=0$, we therefore obtain the useful relation
\begin{equation}
\frac{\dot{\cal R}}{H} = \frac{\MP^2}{a^2 \Sigma} \partial^2 {\cal R} \quad \text{at } H=0. \label{Ratbounce}
\end{equation}
From this, we learn that the crucial quantity $\dot{\cal R}/H$ is finite when $H$ becomes zero. Moreover, it is small for long-wavelength modes due to the double spatial derivative. This result has the immediate implication that the quadratic action in Eq. (\ref{QuadraticRewritten}) is perfectly finite everywhere and, in particular, at the bounce. We note here that one can also solve the equation of motion for the curvature perturbation perturbatively around the bounce \cite{BKLO14}. The result, written in Fourier space, is that near $H=0$
\begin{equation}
{\cal R}_{k} = c_1 \left( 1- \frac{1}{2}c_s^2 k^2 t^2 + \cdots \right) + c_2 t^3 + \dots\,,
\end{equation}
where $c_1, c_2$ are integration constants. This solution is consistent with \eqref{Ratbounce} above.\\
\indent In co-moving gauge, the cubic action is found to be
\begin{eqnarray}
S^{(3)} = \int \mathrm{d} t \mathrm{d} x^3 &a^3& \left[ \left( 3\MP^2 H^2 - P_{,X} X -4 P_{,XX} X^2 - \frac{4}{3} P_{,XXX} X^3 \right) \alpha^3 -6 \MP^2 H \alpha^2 \dot{\cal R} \right. \nonumber \\ && \left. + \left( -9\MP^2 H^2 + 3 P_{,X} X + 6 P_{,XX} X^2 \right) \alpha^2 {\cal R} + 3 \MP^2 \alpha \dot{\cal R}^2  + 18 \MP^2 H \alpha {\cal R} \dot{\cal R} \right. \nonumber \\ && \left. + \left( \frac{27}{2}\MP^2 H^2 + \frac{9}{2} P -9 P_{,X}X \right) \alpha {\cal R}^2 -\frac{2\MP^2}{a^2} \alpha {\cal R} \partial^2 {\cal R} - \frac{\MP^2}{a^2} \alpha (\partial {\cal R})^2 \right. \nonumber \\ && \left. -9\MP^2 \dot{\cal R}^2 {\cal R} -27 \MP^2 H \dot{\cal R} {\cal R}^2 + \frac{9}{2}\left( -3\MP^2 H^2 + P \right) {\cal R}^3 -\frac{\MP^2}{a^2} {\cal R}^2 \partial^2 {\cal R} - \frac{\MP^2}{a^2} {\cal R} (\partial {\cal R})^2 \right. \nonumber \\ && \left. + \frac{\MP^2}{a^2} \left( 2 {\cal R} \dot{\cal R} -2\alpha \dot{\cal R} + H {\cal R}^2 -2H \alpha {\cal R} + 2H \alpha^2 \right) \partial^2 \beta \right. \nonumber \\ && \left.  + \frac{2\MP^2}{a^2} \left( \dot{\cal R} - H \alpha + H {\cal R} \right) \partial {\cal R} \cdot \partial \beta \right. \nonumber \\ && \left. - \frac{\MP^2}{2a^4}({\cal R}+ \alpha) \left( \beta_{,ij} \beta_{,ij} - \partial^2 \beta \partial^2 \beta \right) - \frac{2\MP^2}{a^4} {\cal R}_{,i} \beta_{,j} \beta_{,ij}  \right] \ .
\end{eqnarray}
Terms proportional to the second order perturbation of the lapse function multiply a constraint, and thus do not appear. Substituting $\alpha = \dot{\cal R}/H,$ integrating by parts, discarding total derivatives and employing the background equations of motion we obtain
\begin{eqnarray}
S^{(3)} = \int \mathrm{d} t \mathrm{d} x^3 &a^3& \left[ \left(- P_{,X} X -4 P_{,XX} X^2 - \frac{4}{3} P_{,XXX} X^3 \right) \left(\frac{\dot{\cal R}}{H}\right)^3 \right. \nonumber \\ && \left. + \left(3 P_{,X} X + 6 P_{,XX} X^2 \right) \left( \frac{\dot{\cal R}}{H} \right)^2 {\cal R}  \right. \nonumber \\ && \left.  -\frac{2\MP^2}{a^2} \frac{\dot{\cal R}}{H} {\cal R} \partial^2 {\cal R} - \frac{\MP^2}{a^2} \frac{\dot{\cal R}}{H} (\partial {\cal R})^2 + \frac{\MP^2}{a^2} {\cal R} (\partial {\cal R})^2   \right. \nonumber \\ && \left. + \frac{\MP^2}{2a^4}(3{\cal R} - \frac{\dot{\cal R}}{H}) \left( \beta_{,ij} \beta_{,ij} - \partial^2 \beta \partial^2 \beta \right) - \frac{2\MP^2}{a^4} (\partial{\cal R}\cdot \partial \beta) \partial^2 \beta  \right] \ .
\label{lamp4}
\end{eqnarray} 
A few comments. We have performed fewer integrations by parts than Seery and Lidsey \cite{SL05} and other authors \cite{GS11}. By doing this, we find that no dangerous--looking $1/c_s^2$ terms appear\footnote{Such dangerous terms can appear by performing integrations by parts of the following form: $\int \frac{\Sigma}{H^2} \alpha^2 {\cal R} = \int \frac{\epsilon}{c_s^2} \alpha^2 {\cal R} =  \int \frac{\mathrm{d}}{\mathrm{d}t}(\frac{1}{H}) \frac{1}{c_s^2}\alpha^2 {\cal R} = - \int \frac{1}{c_s^2 H} \alpha^2 \dot{\cal R} + \cdots,$ where the dots include a ``boundary'' term localised at $c_s^2=0.$ If one were to keep this boundary term, the total action would be manifestly non-singular, but often such terms are dropped, leading to naively singular--looking actions.}.
When $H=0,$ there are again several apparently singular terms, but notice that they all involve powers of $\dot{\cal R}/H,$ which we have shown to be finite at the bounce. We  still have to discuss the behaviour of the shift function $\beta$ at the bounce. The linear constraint (\ref{LinearConstraintN}) ``appears'' as though it might cause $\beta$ to blow up at the bounce. Since $\beta$ drops out entirely from the quadratic action, any singularity in $\beta$ would have gone unnoticed to this order. However, it follows from \eqref{lamp4} that $\beta$ blowing up at the bounce would render the cubic action singular. We can combine the constraint for $\beta$ with the equation of motion for ${\cal R}$ to obtain
\begin{equation}
\frac{\MP^2\dot{H}}{a^2 \Sigma}H \partial^2 \beta = H \frac{\mathrm{d}}{\mathrm{d}t}\left( \frac{\dot{\cal R}}{H} \right) + \left( 3H^2 + H \frac{\dot\Sigma}{\Sigma}\right) \frac{\dot{\cal R}}{H} \ .
\end{equation}
Since $\dot{\cal R}/H$ is momentarily constant when $H=0,$ we find from the relation above that
\begin{equation}
\MP^2 \partial^2 \beta = \frac{a^2 \dot\Sigma}{\dot{H}} \frac{\dot{\cal R}}{H} = \frac{\dot\Sigma}{\Sigma\dot{H}}\partial^2{\cal R} \quad \text{at } H=0.
\end{equation}
There is only one independent perturbation variable for systems of gravity coupled to a single scalar field. This is true because out of the $5$ scalar perturbations of the metric and scalar field, two are eliminated by time and space reparameterizations, and two more are eliminated by the constraints. Hence the perturbation $\beta$ must vanish when ${\cal R}$ does, implying that
\begin{equation}
\MP^2 \beta = \frac{\dot\Sigma}{\Sigma\dot{H}} {\cal R} \quad \text{at } H=0.
\end{equation}
Keeping in mind that $\Sigma = \MP^2 H^2 \frac{\epsilon}{c_s^2},$ we can conclude that $\dot\Sigma = 0$ at the moment of the bounce. Hence, we prove the stronger result that
\begin{equation}
\beta = 0 \quad \text{at } H=0.
\end{equation}
Therefore, we can safely ignore all terms involving $\beta$ in our discussions of the behavior of the physical system at and very close to the bounce. That is, we see that the perturbative analysis is indeed non-singular throughout the bouncing spacetime solution.

The results of this Section definitively answer the second and third important questions raised in the Introduction. That is,

\noindent $\bullet$ {\it Will the $1/H$ terms in the cubic action just be ``apparent'' singularities, or do they signal the breakdown of the perturbative description}?

\noindent $\bullet$ {\it It would appear that the cubic action becomes infinite at the moments when $c_s^2=0$, signaling the breakdown of the effective theory. Is this true--or are these singularities only ``apparent'', disappearing upon careful calculation of the cubic action}?

\noindent The answer to the second question is that terms proportional to $1/H$ in the cubic action are actually explicitly finite and, hence, only ``apparent'' singularities. A careful analysis of the cubic action also reveals that there are, in our context, no terms proportional to $1/c_s^2$. This answers the third question. The apparent $1/c_s^2$ divergences on either side of the NEC violating region simply do not exist. The positive answer to both of these questions means that the effective theory of the bounce cosmology is completely finite, singularity free and trustworthy.


\section{Discussion} \label{sec:discussion}

Our results show that there exist effective theories for a non-singular bouncing cosmology where all perturbation modes that lie within the regime of validity of the theory evolve through the bounce in a controlled manner. This includes, in particular, the modes of observational interest in ekpyrotic cosmology. This is a non-trivial result because, in a flat universe, the existence of a bounce requires the null energy condition to be violated. This can be achieved through a temporary phase of ghost condensation, at the expense of a very brief instability due to an imaginary speed of sound. In previous work with L.~Battarra \cite{BKLO14}, we had shown that this instability is in fact too brief to significantly affect long wavelength modes. On the other hand, the same work had also indicated an increasing amplification of ever shorter modes, specifically those whose wavelengths are smaller than the horizon size at the onset of the bounce -- see Fig.~\ref{fig:bounce5}. Hence, one may worry whether these small wavelength, large amplitude modes could destabilize the background evolution. Through a derivation -- carried out in flat gauge -- of the action up to third order in perturbation theory, we have calculated the scale of strong coupling and shown that it is higher than the background energy density throughout the bounce.  We then show that the problematic modes are so short that they are outside of the range of validity of the classical effective theory and, hence, do not disrupt the bouncing spacetime background. This establishes that the bounce solution is trustworthy.

An important aspect of the calculation is that it reveals a decoupling limit (reminiscent of that in Galileon models \cite{Nicolis:2004qq}), in which the scalar field perturbations decouple from the metric perturbations to the extent that the ghost condensate scale is separated from the Planck scale. It is interesting that this decoupling, which intuitively rests on the notion that over sufficiently short distances the metric may be approximated as being flat, also operates in a bouncing spacetime. Furthermore, we have studied the appearance of inverse powers of the Hubble rate and the speed of sound in calculations of the cubic perturbation action -- see, for example,~\cite{SL05}. Both $H$ and $c_{s}^{2}$  necessarily pass through zero during the evolution of the type of bounces that we are studying. Hence, one may wonder whether this will cause perturbation theory to completely break down. Resorting to co-moving gauge to analyze this problem, we have shown that each inverse power of the Hubble rate gets multiplied by the time derivative of the co-moving curvature perturbation, and that this product remains finite. Furthermore, we demonstrated that the dangerous cubic terms proportional to $1/c_{s}^{2}$ simply do not appear in our effective action. It follows that the perturbative analysis is valid throughout the bounce solution. 

Our results are in line with the non-perturbative numerical treatment of Xue et al.~\cite{XGPS13}, where perturbations were also seen to be little affected by their passage through the bounce. However,  their study employs a model with a ghost field and, hence, is ill-defined at the quantum level. In contrast, Peter et al.~\cite{GLP15} found that for curvature-induced bounces in closed universes -- that is, FLRW metrics with curvature parameter $K=+1$ -- perturbations are strongly affected by the bounce. For example, unacceptably large non-Gaussianities are typically generated. We note, however, that such curvature-dominated bounces are highly tuned because matter, radiation and, in particular, anisotropies scale faster than homogeneous curvature in a contracting universe. This makes such curvature-dominated bounces highly unlikely. Our results demonstrate that for {\it flat} FLRW bounces, which in the context of ekpyrotic cosmologies are natural\footnote{This follows from the fact that the ekpyrotic phase strongly suppresses both homogeneous and anisotropic curvature.}, perturbations are essentially unaffected by the bounce. In order to comment on the issue of non-Gaussianity, we should first discuss the implications of our results for model building.

One of our main findings is that the background and cut-off are further separated in the presence of a negative potential during the bounce\footnote{It would also be interesting to see if the two scales can be further separated in more elaborate models including, for instance, Galileon terms, as in the full super-bounce model \cite{KLO14a}. We leave this question for future work.}. This is noteworthy, since negative potentials are natural in ekpyrotic cosmology \cite{Khoury:2001wf,Khoury:2001bz}. However, with regard to the generation of primordial curvature perturbations, the presence of a negative potential during the bounce suggests a small modification of existing scenarios. During the ekpyrotic phase, curvature perturbations are not amplified \cite{Khoury:2001zk,Lyth:2001pf,Tseng:2012qd,Battarra:2013cha}. However, in the presence of a second scalar field, nearly scale-invariant entropy perturbations may be generated \cite{Finelli:2002we,Notari:2002yc,Lehners:2007ac,Ijjas:2014fja}. Note that such a second spectator field does not affect the background evolution and, hence, does not affect the calculations of the present paper. So far, it has typically been assumed that the entropy perturbations get converted into curvature perturbations in between the end of the ekpyrotic phase and the bounce. Were they to get converted while the ekpyrotic potential still dominates the dynamics, the resulting non-Gaussianities could be unacceptably large \cite{Buchbinder:2007at,Koyama:2007if}. This conversion can, for example, occur via a turn in the scalar field trajectory -- see \cite{Lehners:2006pu,Lehners:2006ir} for a concrete model. However, our results now indicate that the bounce is under better control when the ekpyrotic potential is significant throughout the entire NEC violating phase. Hence it may be more natural for the ekpyrotic phase to lead directly into the bounce, with no intermediate kinetic phase. In that case, the potential would turn off again after the bounce, and the conversion of entropy into curvature fluctuations could occur {\it after} the bounce. In this scenario, all adiabatic modes would remain in their quantum vacuum throughout the contracting phase and would only be negligibly amplified during the bounce. There would be entropy perturbations present during the bounce phase, but with no effect on the bouncing spacetime itself. Then, after the bounce, and perhaps during reheating \cite{Battefeld:2007st}, the entropy perturbations would be converted into curvature perturbations and the universe would eventually reach thermal equilibrium, with the hot big bang phase following. In this case, the non-Gaussianity of the curvature perturbations would also be generated after the bounce. In future work, it will be interesting to see whether any of the predictions, in particular those regarding non-Gaussianities \cite{Lehners:2013cka,Fertig:2013kwa,Fertig:2015ola}, are changed when the conversion of entropy into curvature fluctuations occurs after the bounce, rather than before. This will require a separate study.


\acknowledgments 
M.K.~and B.O.~are supported by the DOE under contract No.~DE-SC0007901. J.L.L.~gratefully acknowledges the support of the European Research Council in the form of the Starting Grant No.~256994 ``StringCosmOS''.

\appendix
\section{Vector and tensor perturbations in a bouncing spacetime}

In the main part of the paper, we focussed on scalar fluctuations, which allowed us to calculate the strong coupling scale of the theories we are interested in. However, in general one has to consider not just scalar fluctuations, but also vector and tensor perturbations (by which we mean perturbations transforming as vectors or tensors from the spatial three-dimensional point of view). In this Appendix, we will analyze the behavior of vector and tensor perturbations in a bouncing spacetime. As we will see, these perturbations are not amplified and, hence, we need not consider them in assessing the validity of bouncing solutions. We will comment on the observational significance of this result below.

Under a change of coordinates $x^\mu \rightarrow x^\mu + \xi^\mu$ the metric changes as
\begin{equation}
g_{\mu\nu} \rightarrow g_{\mu\nu} - \nabla_\mu \xi_\nu - \nabla_\nu \xi_\mu = g_{\mu\nu} - g_{\sigma\nu} \partial_\mu \xi^\sigma - g_{\sigma\mu} \partial_\nu \xi^\sigma - g_{\mu\nu,\sigma} \xi^\sigma\,.
\end{equation}
One can decompose $\xi^\mu$ into scalar (2 degrees of freedom) and vector parts (also 2 degrees of freedom):
\begin{equation}
\xi^\mu = (\xi^0,\xi^i) \qquad {\textrm{with}} \quad \xi^i = \xi^i_T + \xi^{,i} \quad \textrm{where} \quad \partial_i \xi^i_T = 0.
\end{equation}
Now consider a perturbed metric in conformal time $\tau$, where we only write out the vector and tensor perturbations at this point. We find
\begin{equation}
\d s^2 = a(\tau)^2 \left[ -\d\tau^2 + 2 S_i \d\tau dx^i + (\delta_{ij}+F_{i,j} + F_{j,i} + \gamma_{ij}) \d x^i \d x^j\right].
\end{equation}
Here we impose that the vector perturbations are transverse, $\partial_i S^i = 0 = \partial_i F^i$ and the tensor perturbations are both transverse and traceless ${\gamma^i}_{j,i}=0={\gamma^i}_i$. Then, under a change of coordinates (where we are now only interested in the vector part $\xi^i_T$), these perturbations change as follows:
\begin{eqnarray} 
S_i \rightarrow S_i - \delta_{ij}\xi_{T,\tau}^{j} \\ F_i \rightarrow F_i - \delta_{ij}\xi_T^{j} \\ \gamma_{ij} \rightarrow \gamma_{ij}\,.
\end{eqnarray}
The tensor perturbations are immediately gauge invariant, but the vector perturbations are not. However, it is easy to see that there exists a gauge-invariant quantity, namely
\begin{equation}
V_i \equiv S_i - F_{i,\tau}\,.
\end{equation}
To obtain the equations for $V_i,$ the simplest procedure is to use $\xi^i_T$ to fix the gauge such that $F_i=0,$ since we can then just replace $S_i$ with $V_i.$  The perturbed Einstein tensor is given in Fourier space by
\begin{eqnarray}
\delta G_{00} &=& 0 \\ \delta G_{0i} &=& V_i (-2 {\cal H}_{,\tau} - {\cal H}^2) + \frac{1}{2} k^2 V_i \\ \delta G_{ij} &=& -\frac{1}{2}\left[ (V_{i,j\tau} + V_{j,i\tau}) + 2 {\cal H} (V_{i,j} + V_{j,i}) \right] \\ &&  + \gamma_{ij} (-2 {\cal H}_{,\tau} - {\cal H}^2) + \frac{1}{2} \left[ \gamma_{ij,\tau\tau} + 2 {\cal H} \gamma_{ij,\tau} + k^2 \gamma_{ij} \right]\,,
\end{eqnarray}
where ${\cal H} \equiv \frac{a_{,\tau}}{a}.$ For a $P(X,\phi)$ theory, the perturbed stress-energy tensor is
\begin{eqnarray}
\delta T_{00} &=& 0 \\ \delta T_{0i} &=& a^2 V_i P \\ \delta T_{ij} &=& a^2 \gamma_{ij} P
\end{eqnarray}

Using the background Einstein equation $2{\cal H}_{,\tau} + {\cal H}^2 + a^2 P =0,$ the linearized equations of motion then become
\begin{eqnarray}
k^2 V_i &=& 0 \\ (V_{i,j\tau} + V_{j,i\tau}) + 2 {\cal H} (V_{i,j} + V_{j,i}) &=& 0 \\  \gamma_{ij,\tau\tau} + 2 {\cal H} \gamma_{ij,\tau} + k^2 \gamma_{ij}  &=& 0.
\end{eqnarray}
Thus, there is no source for either the vector or tensor perturbations. The first equation above then implies that we have no vector perturbations to worry about. Even if there were initial vector perturbations, according to the second equation they would scale as $V_i \propto 1/a^2.$ Thus they cannot compete with the ekpyrotic background (which scales as $\rho \propto  a^{-2\epsilon}$ with $\epsilon > 3$), and would also do very little during a non-singular bounce, as the scale factor evolves very little during the bounce. For long-wavelength modes -- that is, ignoring the $k^2$ term -- the tensor equation above has two solutions; either $\gamma = constant $ or $\gamma\propto 1/a^2$. Again both are harmless. Short-wavelength tensor fluctuations (large $k$) simply oscillate but are not amplified. Note also that they always propagate at the speed of light, and thus, in contrast to the scalar modes, they do not suffer from any gradient-type instability near the bounce. Thus our flat cosmological bounce does not generate any vector or tensor perturbations, nor does it amplify any pre-existing ones. 

The fact that vector perturbations are not amplified in ekpyrotic models is easy to understand: first note that vector perturbations imply a preferred direction in space. But the ekpyrotic phase renders the universe increasingly isotropic and in doing so it suppresses any existing vector perturbations. As discussed above, no additional vector perturbations are then created during the bounce phase. For tensor perturbations, we have a similar outcome. The growth of tensor perturbations is solely dependent on the behavior of the metric. In inflationary models, for instance, the tensor perturbations are amplified because the background spacetime expands in an accelerated fashion \cite{Starobinsky:1979ty}. In ekpyrotic models, we have a rather different situation: the contraction phase proceeds with a very small Hubble rate -- that is, it is a phase of very slow contraction during which the scalar field rolls down a steep and negative potential. A rough approximation to the background spacetime is in fact simply Minkowski space. This rough approximation immediately explains why tensor modes are not amplified in ekpyrotic models \cite{Boyle:2003km} -- they are not amplified around us in our living rooms either! Rather, during the ekpyrotic phase, at linear order in perturbation theory the tensor modes remain in their quantum vacuum state just like the adiabatic modes \cite{Battarra:2013cha}. Thus, to linear order, the tensor-to-scalar ratio $r$ is simply zero. Once curvature fluctuations have been generated (which, as we have discussed, could occur either before or after the bounce), these scalar fluctuations act as a source for the tensor modes at second order in perturbation theory, leading to a small tensor-to-scalar ratio of $r \approx 10^{-6}$  \cite{Baumann:2007zm}. As we have just discussed, even if this tiny tensor spectrum is produced before the bounce, it will not get amplified by the non-singular bounce. Thus ekpyrotic models combined with non-singular bounces predict that no primordial gravitational waves (nor the associated B-mode polarization of the CMB photons) should be detected by near-future experiments (which will optimistically probe down to values of $r \approx 10^{-3}$). It remains to be seen when our observational technologies will be developed enough to detect the tiny $r$ value implied by all currently known ekpyrotic models.

The conclusion of the present Appendix is that it is enough to look at the behaviour of the scalar perturbation modes in assessing the validity of the effective description of non-singular bounces.

\bibliographystyle{apsrev-title}
\bibliography{StrongCoupling}

\end{document}